\documentclass[11pt]{article}
\usepackage[english]{babel}
\usepackage[utf8x]{inputenc}
\usepackage{tikz}
\usepackage{adjustbox}
\usepackage{caption}
\usepackage{amsfonts}
\usepackage{latexsym,amsmath}
\usepackage{amssymb,array}
\usepackage{calc}
\usepackage{graphics}
\usepackage{longtable}
\usepackage{pdflscape}
\usepackage{xcolor}

\usepackage{authblk}

\usepackage[colorlinks=true]{hyperref}
\hypersetup{urlcolor=blue, citecolor=red}
\makeatletter \oddsidemargin 0in \evensidemargin 0in \textwidth
16cm\textheight 20cm 
\begin{document}
\newcommand{\la}{\lambda}
\newcommand{\eq}{\Leftrightarrow}
\newcommand{\mf}{\mathbf}
\newcommand{\ri}{\Rightarrow}
\newtheorem{t1}{Theorem}[section]
\newtheorem{d1}{Definition}[section]
\newtheorem{n1}{Notation}[section]
\newtheorem{c1}{Corollary}[section]
\newtheorem{l1}{Lemma}[section]
\newtheorem{r1}{Remark}[section]
\newtheorem{re1}{Result}[section]
\newtheorem{e1}{Counterexample}[section]
\newtheorem{p1}{Proposition}[section]
\newtheorem{cn1}{Conclusion}[section]
\renewcommand{\theequation}{\thesection.\arabic{equation}}
\pagenumbering{arabic}

\title{Control Chart for Generalized Weibull Quantiles under Hybrid Censoring}

\author[1]{Amarjit Kundu }
\author[2]{Shovan Chowdhury\footnote{Corresponding author e-mail:meetshovan@gmail.com}}
\author[1]{Bidhan Modok}

\affil[1] {Department of Mathematics, Raigunj University, West Bengal, India}
\affil[2] {Quantitative Methods and Operations Management Area, Indian Institute of Management, Kozhikode, India}

\maketitle
\begin{abstract}
In this article, bootstrap and Shewhart type process control monitoring schemes are proposed for the quantiles of generalized Weibull distribution under hybrid censoring. Monitoring schemes for the quantiles of Weibull, generalized exponential, Rayleigh, and Burr type $X$ distributions for type I, type II and hybrid censoring can be obtained as the special cases of the proposed schemes. The maximum likelihood estimators are derived under hybrid censoring using EM algorithm and the asymptotic properties of the estimators are discussed in order to develop the Shewhart type scheme. The in-control performance of the schemes is examined in a simulation study on the basis of the average run length for different choices of quantiles, false-alarm rates and sample sizes. Behavior of the out-of-control performance of the schemes is studied for several choices of shifts in the parameters of the chosen density function. The proposed monitoring schemes are  illustrated with an example from healthcare and compared with similar schemes under type I and type II censoring. The schemes are found to detect out-of-control signals effectively in terms of frequency and speed both.   
\end{abstract}

{\bf Keywords and Phrases}: Average run length, Control scheme, False alarm rate, Generalized Weibull distribution, Hybrid censoring, Parametric bootstrap, Quantile 
\setcounter{section}{0}
\section{Introduction}
\setcounter{equation}{0}
\hspace*{0.2in} The papers by Padgett and Spurrier (1990) and Nichols and Padgett (2005) argued in favor of monitoring lower quartile of strength distribution over average to confirm the quality of carbon fiber strength. Afterward, control schemes for quantiles have received a lot of attention in the quality and reliability literature (see for example, Erto and Pallotta (2007); Lio and Park (2008, 2010); Lio \emph{et al.} (2014); Erto \emph{et al.} (2015); Chiang \emph{et al.} (2017, 2018); Chowdhury \emph{et al.} (2021)). The monitoring schemes, also known as control charts proposed in these papers are applicable to complete data setting only. In practice, reliability data or equivalently, lifetime data are skewed and censored. Recently Vining \emph{et al.} (2016) emphasized on using censored data in reliability studies as customers expected products and processes to perform with high quality over the entire expected lifetime of the product/process. Most of the available control schemes for censored data monitor mean of a process. There are few papers available in the literature discussed monitoring quantiles of a process using censored data. Haghighi \emph{et al.} (2015) proposed control schemes for the quantiles of the Weibull distribution for type-II censored data, based on the distribution of a pivotal quantity conditioned on ancillary statistics. Wang \emph{et al.} (2018) proposed EWMA and CUSUM schemes for monitoring the lower Weibull quantiles under complete data and type-II censoring using the same approach as used in Haghighi \emph{et al.} (2015). To the best of our knowledge, no monitoring scheme is proposed in the literature so far for the hybrid censored data. \\
\hspace*{0.2in} In type-I censoring scheme, a life testing experiment is aborted after a pre-decided time $x_{0}$; whereas in type-II censoring, the termination is subject to failure of a pre-fixed number of items $r$. In case of Type-I censoring, there is a possibility that very few or no failures are observed within the pre-decided time, whereas in Type-II censoring, the experiment can take exceedingly high time to get atleast one failure for not having any upper bound on the test time. In order to address the limitations of the conventional censoring schemes, Epstein (1954) introduced a mixture of Type-I and Type-II censoring schemes, known as type I hybrid censoring. It can be described briefly as follows: Suppose $n$ identical and independent units are put on an experiment. Now if $X_{1:n},...,X_{n:n}$ be the ordered lifetimes of the units, then the experiment is aborted either when a pre-chosen number $r~(<n)$ out of $n$ items has failed or when a pre-determined time $x_{0}$ has elapsed. Hence the life test can be terminated at a random time $X^{*}=min\{X_{r:n},x_{0}\}$. One of the following two types of observations can be witnessed under type-I hybrid censoring scheme. A schematic illustration of the type I hybrid censoring technique is provided in Figure 1. \\
\\
Case I: \{$X_{1:n}<...<X_{r:n}$\} if $X_{r:n}<x_{0}$.\\
Case II: \{$X_{1:n}<...<X_{d:n}<x_{0}$\} if $0\leq d<r$ and $X_{d:n}<x_{0}\leq X_{d+1:n}$.\\

\hspace*{0.2in} In reliability and survival studies, two-parameter Weibull is the most commonly used distribution in both academia and practice. It is generally adequate for modeling monotone hazard rate and is inappropriate for unimodal and bath-tub shaped failure rates. The generalized Weibull (GW) distribution was proposed by Mudholkar and Srivastava (1993) by exponentiating the two-parameter Weibull distribution and was later analyzed extensively by Mudholkar \emph{et al.} (1995, 1996). GW distribution exhibits non-monotone failure rates including bathtub shaped hazard rate and is shown to fit many real life situations better than the conventional exponential, generalized exponential, Weibull or gamma distributions (see Nadarajah \emph{et al.} (2013)). Due to its flexibility and wide scale applicability, GW is chosen as the underlying distribution for the intended purpose. \\


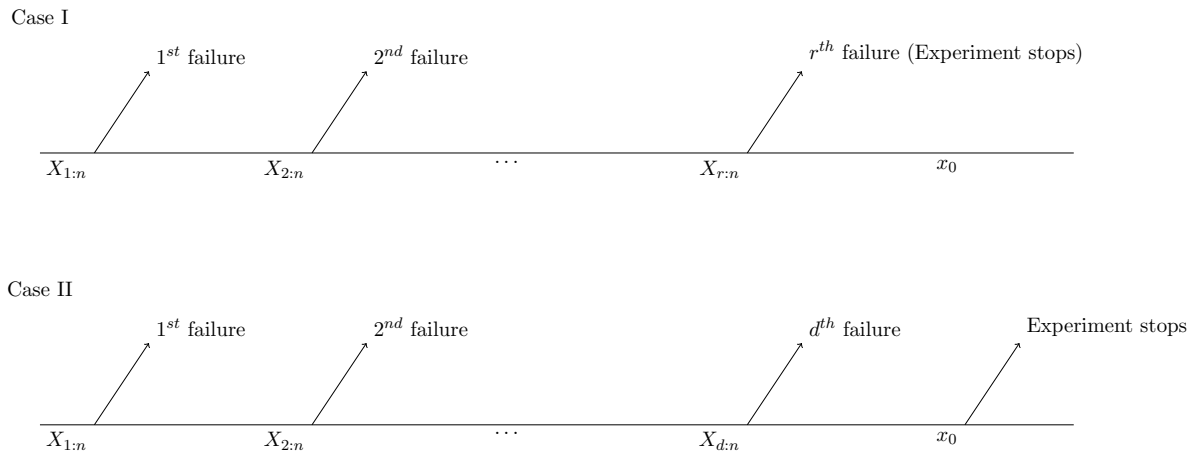
\captionof{figure}{Schematic illustration of type-I hybrid censoring technique}
\begin{center}
\begin{adjustbox}{width=\textwidth}
\begin{tikzpicture}
\draw (1,2.5) node{Case I};
\draw (1,0) -- (20,0);
\draw [->] (2,0) node[anchor=north east] {$X_{1:n}$} -- (3,1.5) node[anchor=south west]{$1^{st}$ failure};
\draw [->] (6,0) node[anchor=north east] {$X_{2:n}$} -- (7,1.5) node[anchor=south west]{$2^{nd}$ failure};
\draw [->] (14,0) node[anchor=north east] {$X_{r:n}$} -- (15,1.5) node[anchor=south west]{$r^{th}$ failure (Experiment stops)};
\draw (18,0) node[anchor=north east] {$x_{0}$};
\draw (10,0) node[anchor=north east]{\ldots};
\draw (1,-2.5) node{Case II};
\draw (1,-5) -- (20,-5);
\draw [->] (2,-5) node[anchor=north east] {$X_{1:n}$} -- (3,-3.5) node[anchor=south west]{$1^{st}$ failure};
\draw [->] (6,-5) node[anchor=north east] {$X_{2:n}$} -- (7,-3.5) node[anchor=south west]{$2^{nd}$ failure};
\draw [->] (14,-5) node[anchor=north east] {$X_{d:n}$} -- (15,-3.5) node[anchor=south west]{$d^{th}$ failure};
\draw [->] (18,-5) node[anchor=north east] {$x_{0}$} -- (19,-3.5) node [anchor=south west]{Experiment stops};
\draw (10,-5) node[anchor=north east]{\ldots};
\end{tikzpicture}
\end{adjustbox}
\end{center}

\hspace*{0.2in} The monitoring schemes require estimation of parameters associated with the underlying distribution. Research papers on the classical estimation under hybrid censoring scheme are available in the literature for many important and popular distributions. For a detailed discussion on the inferential results under hybrid censoring, one can refer to Balakrishnan and Kundu (2013) and the references therein. In the present work, the parameters of GW distribution are estimated using maximum likelihood (ML) method and bootstrap method. The ML estimators of the parameters of GW distribution under hybrid censoring have not been studied before. The estimators do not have any closed form expression and are computed using EM algorithm. The Fisher information matrix, obtained using the missing information principle (Louise, 1982) and the asymptotic properties of the estimators are used to develop the Shewhart type scheme. Due to the non-availability of closed form expressions of the sampling distribution of the quantiles, parametric bootstrap method is also used to obtain the control limits. For more  discussion on the bootstrap technique and its advantages, one can refer to Efron and Tibshirani (1993), Liu and Tang (1996), Seppala \emph{et al.} (1996) and Jones and Woodall (1998). To the best of our knowledge, no study has been found on developing monitoring schemes for quantiles under hybrid censored GW distributed data. Moreover, the control schemes for quantiles of Weibull, generalized exponential, Rayleigh, and Burr type $X$ distributions for type I, type II and hybrid censoring can be obtained as the special cases of the schemes proposed in this article.\\  

\hspace*{0.2in} The rest of the paper is organized as follows. Section $2$ provides Fisher information matrix and asymptotic results of the ML estimators required to develop Shewhart type scheme. The proposed bootstrap and Shewhart-type schemes for GW quantiles are introduced in Section $3$. Section $4$ is devoted to the practical implementation of the scheme including tabulation of the control limits and average run length (ARL). Simulation results of both in-control (IC) and out-of-control (OOC) performance of the bootstrap scheme are presented in Section $4$. The effectiveness of the proposed schemes is evaluated in Section $5$ using a skewed data set from healthcare. Section $7$ concludes the paper.

\setcounter{section}{1}
\section{Statistical Framework}
\setcounter{equation}{0} In this section, MLEs of the parameters of GW distribution under hybrid censoring are obtained using EM algorithm followed by a discussion on the Fisher information matrix and asymptotic properties of the estimators. 
\subsection{Underlying Distribution}
\hspace*{0.2in} Let $X$ be a random variable following three parameters GW distribution with scale parameter $\lambda>0$, the shape parameters $\alpha>0$ and $\theta>0.$ Then probability density function (pdf) and cumulative distribution function (cdf) of $X$ are given by  
\begin{equation}\label{e1}
f\left(x| \theta,\alpha,\lambda\right)=\lambda \alpha \theta  (1-e^{-\lambda x^\theta})^{\alpha -1} e^{-\lambda x^\theta}x^{\theta-1};~x>0,
\end{equation}\
and 
\begin{equation}\label{e2}
F\left(x|\theta,\alpha\right)=\left(1-e^{-\lambda x^\theta}\right)^{\alpha}.
\end{equation}
The following pdfs are obtained as the special cases of GW distribution.
\begin{itemize}
\item Weibull distribution for $\theta=1,$
\item Generalized exponential for $\alpha=1,$
\item Burr type $X$ for $\alpha=2,$
 \item Rayleigh for $\alpha=2$ and $\theta=1.$
\end{itemize}
It is well known that the hazard function of GW distribution allows for both monotone and non-monotone failure rates. In particular, GW distribution is bathtub shaped when $\alpha>1$ and $\alpha\theta<1,$ unimodal when $\alpha<1$ and $\alpha\theta>1,$ monotonically increasing (decreasing) when $\alpha> (<)~1$ and $\alpha\theta> (<)~1.$ The hazard function of the distribution is independent of the scale parameter $\lambda.$ The assumption of $\lambda=1$ is used throughout the paper.\\
Let $\xi_p$ be the $p^{th}$ quantile of the GW distribution which can be obtained by the following equation
\begin{equation}\label{e3}
\xi_p=F^{-1}(p;\theta,\alpha)=\left[\ln\left(\frac{1}{1-p^{1/\alpha}}\right)\right]^{1/\theta}.
\end{equation}

\subsection{Maximum Likelihood Estimators}
\hspace*{0.2in} Let $x_{i_1},x_{i_2},\ldots x_{i_n}$ be $i^{th}$ in-control (IC) random subgroup of size $n$ $(i=1,2,\cdots,k)$ drawn from phase I process following GW distribution as in (\ref{e1}). On the  basis of the observed data and ignoring the additive constant, the log-likelihood function under hybrid censoring (for Case I and II as introduced in Section 1) is given by 
\begin{eqnarray}\label{e4}
L\left(\theta,\alpha|data\right)&=& d\ln\alpha+d\ln \theta+\left(\theta-1\right)\sum_{i=1}^{d}\ln x_{i:n}-\sum_{i=1}^{d}{x_{i:n}^{\theta}}+\left(\alpha-1\right)\sum_{i=1}^{d}{\ln(1-e^{-x_{i:n}^{\theta}})}\nonumber\\
&+&(n-d)\ln\left[\left(1-\left(1-e^{-c^{\theta}}\right)^{\alpha}\right)\right],
\end{eqnarray}
where for Case I, $d=r$ and $c=x_{r:n}$, and for Case II, $0\leq d\leq r-1$ and $c=x_0.$\\
The MLEs $\hat\theta$ and $\hat\alpha$ are obtained by maximizing the log-likelihood function (\ref{e4}), and subsequently solving the non-linear equations$$ \frac{\partial L}{\partial \theta}=0, \, \frac{\partial L}{\partial \alpha}=0.$$ As closed-form solutions of these two equations are not available, we intend to use EM algorithm to find MLE of the parameters of GW distribution. Let us define ${\bf x}=(x_{1:n},x_{2:n},\cdots, x_{d:n})$ and ${\bf y}=(y_1,y_2, \cdots ,y_{n-d})$ as the observed and censored data respectively with ${\bf w}=({\bf x},{\bf y})$ as the complete data. So, the log-likelihood function of the complete data set is given as
\begin{eqnarray}\label{e5}
L_c\left(w;\theta,\alpha\right)&=&n\ln\alpha+n\ln\theta+(\theta-1)\left[\sum_{i=1}^d{\ln x_{i:n}}+\sum_{i=1}^{n-d}{\ln y_i}\right]
                               -\left[\sum_{i=1}^d{x_{i:n}^{\theta}}+\sum_{i=1}^{n-d}{y_i}^{\theta}\right]\nonumber\\
															&+&(\alpha-1)\left[\sum_{i=1}^d{\ln(1-e^{-x_{i:n}^{\theta}})}+\sum_{i=1}^{n-d}\ln(1-e^{-y_i^{\theta}})\right].
\end{eqnarray}
Now, for the `E'-th step of the EM algorithm, we have to introduce the complete pseudo log-likelihood function as $L_s\left(\theta,\alpha | \text{data}\right)=E\left(L_c\left(W;\theta,\alpha \right)|X\right)$. So, 
\begin{eqnarray}\label{e6}
L_s\left(\theta,\alpha | \text{data}\right) &=& n\ln\alpha+n\ln\theta+(\theta-1)\sum_{i=1}^d{\ln(x_{i:n})}-\sum_{i=1}^d{x_{i:n}^{\theta}}+(\alpha-1)\sum_{i=1}^d{\ln(1-e^{-x_{i:n}^{\theta}})}+(\theta-1)\nonumber\\ 
&&\sum_{i=1}^{n-d}{E\left[\ln Y_i|Y_i>c\right]}-\sum_{i=1}^{n-d}{E\left[Y_i^{\theta}|Y_i>c\right]}+(\alpha-1)\sum_{i=1}^{n-d}E\left[\ln(1-e^{-Y_i^{\theta}})|Y_i>c\right].
\end{eqnarray}
Let $f_{GW}$ and $F_{GW}$ denote the pdf and cdf of GW distribution respectively. The following results are required to calculate the expectations in (\ref{e6}). The proofs of the results are similar to Ng \emph{et al.} $(2002)$, and hence are omitted.
\begin{re1}\label{r1}
Given $X_{1:n}=x_{1:n},X_{2:n}=x_{2:n}, \cdots,X_{r:n}=x_{r:n},$ the conditional pdf of $y_j,$ for $j=1,2,\cdots, n-r$ is 
$$f_{y|x}\left(y_j|x_{1:n},x_{2:n}\cdots,x_{r:n}\right)=\frac{f_{GW}\left(y_j;\theta,\alpha\right)}{1-F_{GW}\left(x_{r:n},\theta,\alpha,\right)}, \; y_j>x_{r:n},$$ 
and $y_j$ and $y_k$ for $j\neq k$ are conditionally independent.
\end{re1}

\begin{re1}\label{r2}
Given $X_{1:n}=x_{1:n},X_{2:n}=x_{2:n}, \cdots,X_{d:n}=x_{d:n}<x_0,$ the conditional pdf of $y_j,$ for $j=1,2,\cdots, n-d$ is  
$$f_{y|x}\left(y_j|x_{1:n},x_{2:n},\cdots,x_{d:n}<x_0\right)=\frac{f_{GW}\left(y_j;\theta,\alpha\right)}{1-F_{GW}\left(x_0;\theta,\alpha\right)}, \; y_j>x_0,$$ 
and $y_j$ and $y_k$ for $j\neq k$ are conditionally independent.
\end{re1}
Assuming
\begin{eqnarray*}
 A\left(c,\theta,\alpha\right)=E\left[\ln Y_i|Y_i>c\right]&=&\int_{c}^{\infty}{\frac{\ln x  f_{\text{GW}}(x)}{1-F_{GW}(c)}dx}\\
                            &=&\frac{\alpha}{1-F_{GW}(c)}\int_{1-e^{-c^{\theta}}}^1{\ln\left(\ln(1-t)^{-1}\right)t^{\alpha-1} dt}\nonumber,
\end{eqnarray*}
\begin{eqnarray*}
 B\left(c,\theta,\alpha\right)=E\left[Y_i^{\theta}|Y_i>c\right]&=&\int_c^{\infty}{x^{\theta}\frac{f_{\text{GW}}(x)}{1-F_{\text{GW}}(c)}dx}\\
																	&=&-\frac{\alpha}{1-F_{\text{GW}}(c)}\int_{1-e^{-c^{\theta}}}^1{t^{\alpha-1}\ln (1-t)dt},
\end{eqnarray*}
and
\begin{eqnarray*}
 C\left(c,\theta,\alpha\right)=E\left[\ln(1-e^{-y_i^{\theta}})|Y_i>c\right]&=& \frac{1}{1-F_{\text{GW}}(c)}\int_c^{\infty}{\ln(1-e^{-x^{\theta}})f_{\text{GW}}(x)dx}\\
																						&=& \frac{\alpha}{1-F_{\text{GW}}(c)}\int_{1-e^{-c^{\theta}}}^1 {t^{\alpha-1}\ln t dt}.
\end{eqnarray*}
the expression in (\ref{e6}) can be written as
\begin{eqnarray}\label{e7}
L_s\left(\theta,\alpha | \text{data}\right) &=& n\ln\alpha+n\ln\theta+(\theta-1)\sum_{i=1}^d{\ln(x_{i:n})}-\sum_{i=1}^d{x_{i:n}^{\theta}}+(\alpha-1)\sum_{i=1}^d{\ln(1-e^{-x_{i:n}^{\theta}})}\nonumber\\ 
&+&(\theta-1) (n-d)A\left(c,\theta,\alpha\right)-(n-d)B\left(c,\theta,\alpha\right)\nonumber\\
&+&(\alpha-1)(n-d)C\left(c,\theta,\alpha\right).
\end{eqnarray}
Now the 'M'-step involves maximization of pseudo log-likelihood function given by (\ref{e7}). The maximization is carried out by solving the equations $\frac{\partial L_s\left(\theta,\alpha | \text{data}\right)}{\partial\theta}=0$ and $\frac{\partial L_s\left(\theta,\alpha | \text{data}\right)}{\partial\alpha}=0$ recursively using Newton-Raphson method. The MLEs once obtained using EM algorithm can be used to estimate the $p^{th}$ quantile, denoted by $\hat{\xi_p}$, and is obtained as
\begin{equation}\label{e8}
\hat{\xi_p}=\left[\ln\left(\frac{1}{1-p^{1/\hat{\alpha}}}\right)\right]^{1/\hat{\theta}}.
\end{equation}

\subsection{Asymptotic Properties}
\hspace*{0.2in} The Fisher information matrix and asymptotic properties of the estimators are discussed here. Using the missing value principle of Louise (1982) it can be written that 
\begin{equation}\label{e10}
\text{Observed information = Complete information - Missing information,}
\end{equation}
and can be expressed as 
\begin{equation}\label{e11}
I_X({\bf\Theta}) = I_{W}\left({\bf\Theta}\right)-I_{W|X}({\bf\Theta}),
\end{equation}
where ${\bf\Theta} \,= \,\left(\theta, \alpha\right)$, and $I_W({\bf\Theta})$ and $I_{W|X}({\bf\Theta})$ denote the complete and missing information respectively. The complete information $I_W({\bf\Theta})$ is given by $$I_W({\bf\Theta}) = -E\left[\frac{\partial^2L_c\left(W;{\bf\Theta}\right)}{\partial{\bf\Theta}^2}\right],$$ with the Fisher information matrix of the censored observations being written as $$I_{W|X}({\bf\Theta})=-(n-d)E_{Y|X}\left[\frac{\partial^2 \ln f_Y(y|X, {\bf\Theta})}{\partial{\bf\Theta}^2}\right].$$ The asymptotic variance covariance matrix of $\hat\Theta$ can be obtained by inverting $I_X\left(\hat{{\bf\Theta}}\right)$. Now, if $$I_W({\bf\Theta})=\begin{bmatrix}
a_{11}(c;\theta,\alpha) & a_{12}(c;\theta,\alpha)\\
a_{21}(c;\theta,\alpha) & a_{22}(c;\theta,\alpha)
\end{bmatrix}, $$  
then all the elements of the above matrix can be computed as
\begin{eqnarray*}
a_{11}=-E\left(\frac{\partial^2L_c\left(W;{\bf\Theta}\right)}{\partial\theta^2}\right)&=&\frac{n}{\theta^2}-\frac{n\alpha}{\theta^2}\int_0^1{\ln z\left(\ln \left(-\ln z\right)\right)^2\left(1-z\right)^{\alpha-1}dz},\\
&+&\frac{n(\alpha-1)\alpha}{\theta^2}\int_0^1{z(1-z)^{\alpha-3}\left(1+\ln z-z\right)\ln z\left(\ln\left(-\ln z\right)\right)^2}dz,\\
a_{22}=-E\left(\frac{\partial^2L_c\left(W;{\bf\Theta}\right)}{\partial\alpha^2}\right)&=&\frac{n}{\alpha^2},\\
\end{eqnarray*}
and
$$a_{12}=a_{21}=-E\left(\frac{\partial^2L_c\left(W;{\bf\Theta}\right)}{\partial\theta\partial\alpha}\right)=\frac{n\alpha}{\theta}\int_0^1{z\ln z\ln\left(-\ln z\right)(1-z)^{\alpha-2}}dz.$$ 
 Again if $$I_{W|X}({\bf\Theta})=(n-d)\begin{bmatrix}
b_{11}(c;\theta,\alpha) & b_{12}(c;\theta,\alpha)\\
b_{21}(c;\theta,\alpha) & b_{22}(c;\theta,\alpha)
\end{bmatrix}, $$
then all the entries of the above matrix can be computed as

\begin{eqnarray*}
b_{11} &=& -E\left(\frac{\partial^2\ln f_{Y|X}(y|X, {\bf\Theta})}{\partial\theta^2}\right)\\
&=& \frac{1}{\theta^2}+c^{\theta}\alpha \ln c\left\{ \frac{\ln c}{\left(z_1^{-\alpha}-1\right)}\left(1-\frac{1}{z_1}\right)+\frac{dz_1}{d\theta}\frac{1}{\left(z_1^{-\alpha}-1\right)^2}\left(\alpha z_1^{-\alpha-1}+\frac{(1-\alpha)z_1^{-\alpha}-1}{z_1^2}\right)\right\}\\
&-& \frac{\alpha}{\theta^2}\int_0^1{\ln z \left(\ln \left(-\ln z\right)\right)^2\left\{(1-z)^2-(\alpha-1)z\ln z-(\alpha-1)z(1-z)\right\}(1-z)^{(\alpha-3)}dz},\\
b_{22} &=& -E\left(\frac{\partial^2\ln f_{Y|X}(y|X, {\bf\Theta})}{\partial\alpha^2}\right) = \frac{1}{\alpha^2}-\frac{(\ln z_1)^2z_1^{\alpha}}{\left(z_1^{-\alpha}-1\right)^2},
\end{eqnarray*}

and
\begin{eqnarray*}
b_{12}=b_{21} &=& -E\left(\frac{\partial^2\ln f_{Y|X}(y|X,{\bf\Theta})}{\partial\theta\partial\alpha}\right)
	= \frac{\alpha}{\theta}\int_0^1{\ln \left(-\ln z\right) z\ln z(1-z)^{(\alpha-2)}dz}\\
&+& c^{\theta} \ln c\left\{ \frac{1}{z_1^{-\alpha}-1}-\frac{1}{z_1^{1-\alpha}-z_1}\right\}+\alpha c^{\theta}\ln c \left\{\frac{z_1^{-\alpha}\ln z_1}{\left(z_1^{-\alpha}-1\right)^2}-\frac{\left(z_1^{1-\alpha} \ln z_1\right)+1}{\left(z_1^{1-\alpha}-z_1\right)^2}\right\}.
\end{eqnarray*}
where $z_1=\left(1-e^{-c^{\theta}}\right).$\\

For $\theta, \alpha>0$, the GW distribution satisfies all the regularity conditions of MLEs (see Mudholkar \emph{et al.} (1996)). Therefore, the result below easily follows from the standard asymptotic distribution results of the MLEs $$\sqrt{n}\left(\hat{\bf\Theta}_n-{\bf\Theta}\right)\rightarrow N_2\left({\bf 0},I^{-1}_X({\bf\Theta})\right).$$
Suppose $\hat{\xi}_p\left(\hat{{\bf\Theta}}_n\right)$ be the value of $\xi_p$ at ${\bf\Theta}=\hat{{\bf\Theta}}_n,$ obtained from $(\ref{e3})$ and calculated on the basis of $n$ observations. It can be shown that $\hat\xi_p\left(\hat{\bf\Theta}_n\right)$ follows asymptotic normal distribution with mean $\xi_p\left({\bf\Theta}\right)$ and variance $\frac{1}{n}{\bf\nabla}\xi_p^{T}\left({\bf \Theta}\right) I^{-1}_{X}\left({\bf\Theta}\right){\bf\nabla}\xi_p\left({\bf \Theta}\right)$, where ${\bf\nabla}\xi_p\left(\Theta\right)$ is the gradient of $\xi_p\left(\Theta\right)$ with respect to $\Theta$ with $\xi_p^{T}\left({\bf \Theta}\right)$ being its transpose. In practice, $I_X ({\bf\Theta})$ is replaced by the observed Fisher Information matrix $\hat I_X\left(\hat{\bf\Theta}_n\right)$, obtained by substituting the unknown parameters $\theta$ and $\alpha$ by their respective MLEs.

\setcounter{section}{2}
\section{Construction of Proposed Control Schemes}
\setcounter{equation}{0} Here we discuss the step by step procedures for developing the bootstrap and Shewhart type process control schemes for quantiles of hybrid censored GW distributed data. The names 'bootstrap hybrid-censored (BHC)' and 'Shewhart-type hybrid-censored (SHC)' will be used hereafter to indicate bootstrap based and Shewhart based monitoring schemes.    

\subsection{Charting Procedure for BHC Scheme}
\hspace*{0.2in} Here, the BHC scheme for the GW quantiles is proposed using the following charting procedure.
\begin{enumerate}
\item[{\bf Step-1:}] Collect and establish $k$ reference samples $X_n=(x_{i1},x_{i2},\ldots,x_{in})$ of size $n$ each from an IC process (Phase I process) following GW cdf $F(x|\theta,\alpha )$ as in (\ref{e2}).
\item[{\bf Step-2:}] Obtain the MLEs of $\theta$ and $\alpha$ from Step-$1$ under hybrid censoring following the procedure described in Section 2.2 and estimate the cdf as $F(x|\hat\theta,\hat\alpha)$.
\item[{\bf Step-3:}] Generate a bootstrap sample of size $m$, $x_1^*,x_2^*,\ldots,x_m^*,$ from $F(x|\hat\theta,\hat\alpha)$ as obtained in Step-2.
\item[{\bf Step-4:}] Obtain the MLEs of $\theta$ and $\alpha$ under hybrid censoring using the bootstrap sample obtained in Step-$3,$ and denote these as $\theta^*$ and $\alpha^*.$
\item[{\bf Step-5:}] Using $(\ref{e3})$ compute the bootstrap estimate of the $p^{th}$ quantile as
\begin{equation} \label{e13}
\hat\xi_p^*=F^{-1}(p; \hat\theta^*,\hat\alpha^*).
\end{equation}
\item[{\bf Step-6:}] Repeat Steps $3$-$5$ large number of times ($B$, say) to obtain bootstrap estimates of $\hat\xi_p^*$, denoted by $\hat\xi_{1p}^*,\hat\xi_{2p}^*,\ldots,\hat\xi_{Bp}^*.$
\item[{\bf Step-7:}] Using $B$ bootstrap estimates as obtained in Step $6$, find the $\frac{\nu}{2}^{th}$ and $(1- \frac{\nu}{2})^{th}$ empirical quantiles as the lower control limit ($LCL$) and upper control limit ($UCL$) respectively to construct a two-sided BHC scheme, where $\nu$ is the false alarm rate (FAR) defined as the probability that an observation is considered as  out of control (OOC) when the process is actually IC. Here, empirical sample quantiles are obtained following a method proposed by Hyndman and Fan (1996).
\item[{\bf Step-8:}] Sequentially observe the $j^{th}$ phase II (test) sample $Y_{j:m}=(Y_{j1},Y_{j2},\ldots,Y_{jm})$ of size $m, j=1,2,…$.
\item[{\bf Step-9:}] Sequentially obtain $\hat{\xi_{jp}}$ using (\ref{e13}) after obtaining MLEs of the parameters using the $j^{th}$ test sample as described in Step-$5.$   
\item[{\bf Step-10:}] Plot $\hat{\xi_{jp}}$ against $LCL$ and $UCL$ as obtained in Step-$7$ of the Phase I process.
\item[{\bf Step-11:}] If $\hat{\xi_{jp}}$ falls in between the $LCL$ and $UCL,$ then the process is assumed to be in-control, otherwise, an OOC signal is activated.
\end{enumerate}

\subsection{Charting Procedure for SHC Scheme}
\hspace*{0.2in} Shewhart-type scheme for the quantiles of GW distribution is developed in this section following the asymptotic properties of the MLEs obtained in Section 2.3. The steps for designing the SHC scheme for $p^{th}$ quantile, $\xi_p\left(\Theta\right)$, are described as follows:\\
In phase I, samples are drawn from in-control process following GW distribution in $k$ independent random subgroups of size $m$ each with $n=m\times k$ being the total sample size. 
\begin{enumerate}
\item[\bf Step-1:] Following the method as described in Section $2.2,$ the MLEs $\hat\alpha_n=\left(\hat\theta_n,\hat\alpha_n\right)$ are computed on the basis of $n$ in-control sample values of Phase I process. Then the asymptotic standard error of $\hat\xi_{p,m}\left(\hat\Theta_m\right)$ is computed as
\begin{equation}\label{e14}
SE_{\xi_{p,m}}=\sqrt{\frac{1}{m}{\bf\nabla}\xi_p^{T}\left(\hat{\bf\Theta}_n\right){\bf I}_n^{-1}\left(\hat{\bf\Theta}_n\right){\bf\nabla}\xi_p\left(\hat{\bf\Theta}_n\right)},
\end{equation}
 where ${\bf\nabla}\xi_p\left(\hat{\bf\Theta}_n\right)$ is the gradient of $\xi_p\left({\bf\Theta}\right)$ at ${\bf\Theta}=\hat{\bf\Theta}_n$ and $I_n({\bf\Theta})$ is the Fisher information matrix of the observed values to be calculated by the method as describe in Section $2.3.$\\
\item[{\bf Step-2:}] The MLEs $\hat{\bf\Theta}^j_m$  of $\Theta$ and $\xi_p^j\left(\hat{\bf\Theta}^j_m\right)$ are calculated based on $j^{th}$ ($j=1,2,\ldots, k$) IC samples of size $m$ each. The sample mean of $\xi_p^j\left(\hat{\bf\Theta}^j_m\right)$s is calculated as 
\begin{equation}\label{15}
\bar\xi_p\left(\hat{\bf\Theta}_m\right)=\frac{1}{k}\sum_{i=1}^{k}\xi_p^j\left(\hat{\bf\Theta}^j_m\right).
\end{equation}
\item[{\bf Step-3:}] The Shewhart-type scheme has the center line $CL_{SH}=\bar\xi_p\left(\hat{\bf\Theta}_m\right)$. If $\nu$ is the false alarm rate (FAR), then for $0<\nu<1$, the upper and lower control limits of the SHC scheme are found to be 
$$UCL_{SH}=\bar\xi_p\left(\hat{\bf\Theta}_m\right)+z_{\left(1-\nu/2\right)}SE_{\xi_{p,m}},$$
and 
 $$LCL_{SH}=\bar\xi_p\left(\hat{\bf\Theta}_m\right)-z_{\left(1-\nu/2\right)}SE_{\xi_{p,m}}$$
respectively, where $z_{\left(1-\nu/2\right)}$ is the $1-\nu/2$th quantile of the standard normal distribution.
\end{enumerate}

\setcounter{section}{3}
\section{Simulation Study}
\setcounter{equation}{0} A comprehensive simulation is carried out in this section to evaluate the IC and OOC performances of the proposed BHC monitoring scheme. Numerical computation in $R$ (version $4.0.2$) based on Monte-Carlo simulations is used to determine the average $UCL$ and $LCL.$ The MLEs of the parameters $\theta$ and $\alpha$ are obtained for the pair $(\theta=0.51, \; \alpha=11.1)$. The control limits are obtained based on $B = 5,000$ bootstrap samples. For given subgroup size $k=20,$ the simulations are performed for different bootstrap sample size $m$, quantiles $(p=0.1, \; 0.5, \; 0.9),$ levels of FAR $(\nu=0.005, \; 0.0027, \;0.002)$ and the following censoring schemes: Scheme $1:\; m=25, \; r=15, \,x_0=55$; Scheme $2: \; m=25, \; r=20, \; x_0=55 $; Scheme $3:\; m=40, \; r=30, \;x_0=55$; Scheme $4:\; m=40, \; r=35, \;x_0=55$; Scheme $5:\; m=25, \; r=15, \;x_0=70$; Scheme $6:\; m=25, \; r=20, \;x_0=70$; Scheme $7:\; m=40, \; r=30, \;x_0=70;$ and Scheme $8:\; m=40, \; r=35, \;x_0=70$. The performance of the scheme is assessed by run length, defined as the number of cases required to observe the first OOC signal. For each simulation, the run length is obtained, followed by obtaining the average run length (ARL) and the standard deviation of run length (SDRL) by using $5,000$ simulation runs.

\subsection{IC scheme Performance }
\hspace*{0.2in} The estimated IC control limits of the BHC scheme are displayed in Table \ref{tab1} along with the respective ARL and SDRL as the scheme performance measures, denoted by $ARL_0$ and $SDRL_0$ respectively. It is easy to show that the reciprocal of FAR is same as the nominal (theoretical) ARL, viz. for $\nu=0.005,\;0.0027$and $0.002$  the nominal ARL should be equal to $200,$ $370$ and $500$  respectively. In general, the smaller ARLs indicate narrower control limits, while ARLs larger than $370$ specifies wider limits that the bootstrap control schemes give fewer false signals. The simulated values of $ARL_0$ in Table \ref{tab1} are found to be closer to the theoretical results implying that the BHC scheme for quantiles perform well with skewed data. As the bootstrap sample size $(m)$ increases, the estimated control limits get closer together. Moreover, for fixed $m,$ the control limits become farther apart as the percentile $(p)$ increases. Also, $SDRL_0$ is found to be closer to the $ARL_0,$ satisfying the theoretical result of the geometric distribution used as the run length model.

\subsection{OOC scheme Performance }
\hspace*{0.2in} The OOC performance of the BHC monitoring scheme is investigated by measuring impact of changes in the IC parameter estimates on ARL. In other words, the phase II sample is considered taken from $GW(\theta+\Delta\theta, \; \alpha+\Delta\alpha),$ while the IC sample comes from $GW(\theta, \;\alpha).$ The effects of shifts $(\Delta\theta~\text{and/or}~\Delta\lambda)$ in the parameters of the GW distribution on the ARL of the quantiles scheme is examined and exhibited in Table \ref{tab2}. In general, the simulation results reveal that for fixed $m,$ $r$, and $x_0$, the OOC ARL values (denoted by $ARL_1$) for the quantiles decrease sharply with both downward and upward small and medium shifts in the parameters indicating the effectiveness and usefulness of the scheme. However, the speed of detection varies depending on the type of shifts, the parameters, and the quantiles being considered. In general, the monitoring scheme is highly sensitive to a minor shift in the IC parameters and detects OOC signals faster. Except for minor sampling fluctuations, we find some interesting patterns in the speed of detection. In particular, when $\alpha$ is IC, the scheme detects OOC signal faster for downward shifts than the upward shifts (refer Table \ref{tab2} and Figure \ref{figure1} (b)). The ARLs around $50^{th}~(90^{th}$) quantile are found to be smaller than the other quantiles for downward (upward) shifts in $\theta.$ For example, for a $4\%$  decrease (increase) in $\theta$ when $\alpha$ is IC $(\Delta\alpha=0),$ there is about $77.6\%~(36.3\%)$ reduction in the $ARL$~of the $50^{th}$ quantile. On the other hand, when $\theta$ is IC, the ARLs around $10^{th}~(50^{th}$) quantile is found to be smaller than other quantiles for downward shifts in $\alpha$ (refer Table \ref{tab2} and Figure \ref{figure1}(a)). For example, there is about $42.1\%~(35.9\%)$ reduction in the $ARL$~of the $50^{th}$ quantile for a $8\%$ decrease (increase) in $\alpha$ when $\theta$ is IC. From Table \ref{tab2} and Figure \ref{figure1}(d) it is also clear that for $4\%$ deviation in $\alpha$ the ARLs around $50^{th}~(90^{th}$) quantile are smaller than the other quantiles for downward shifts in $\theta$. Again, from Figure \ref{figure1}(c) it can also be observed that, for $4\%$ deviation in $\theta$ the ARLs around $90^{th}$ quantile are smaller than the other quantiles irrespective of the direction of the shifts in $\alpha$.

\setcounter{section}{4}
\section{Illustrative Example}
\setcounter{equation}{0}
\hspace*{0.2in} In this section, one numerical example from Lee and Wang (2003) is used to illustrate the BHC and SHC monitoring schemes. The bootstrap schemes with Type-I and Type-II censoring are also illustrated with the same example. The data set describes remission times (in months) of a random sample of 125 bladder cancer patients. The summary measures of the data set are shown below. 
\begin{center}
\[
\begin{array} {ccccccccc}
 Min     & \hspace*{0.1 in}5\%  &\hspace*{0.1 in}  10\%    & \hspace*{0.1 in}  25\%   &\hspace*{0.1 in}  50\%  &\hspace*{0.1 in}    75\%  &\hspace*{0.1 in}  90\%  &\hspace*{0.1 in}   95\%   &\hspace*{0.1 in}  Max\\

0.200 & \hspace*{0.1 in} 1.078  &\hspace*{0.1 in} 1.864   &\hspace*{0.1 in} 3.360  &\hspace*{0.1 in}  6.760  &\hspace*{0.1 in}   11.980 &\hspace*{0.1 in}  19.820 &\hspace*{0.1 in}  26.212  &\hspace*{0.1 in} 79.050
\end{array}
\]
\end{center}

\begin{figure}[ht]
\centering
\includegraphics[height=7.5 cm,width=15 cm]{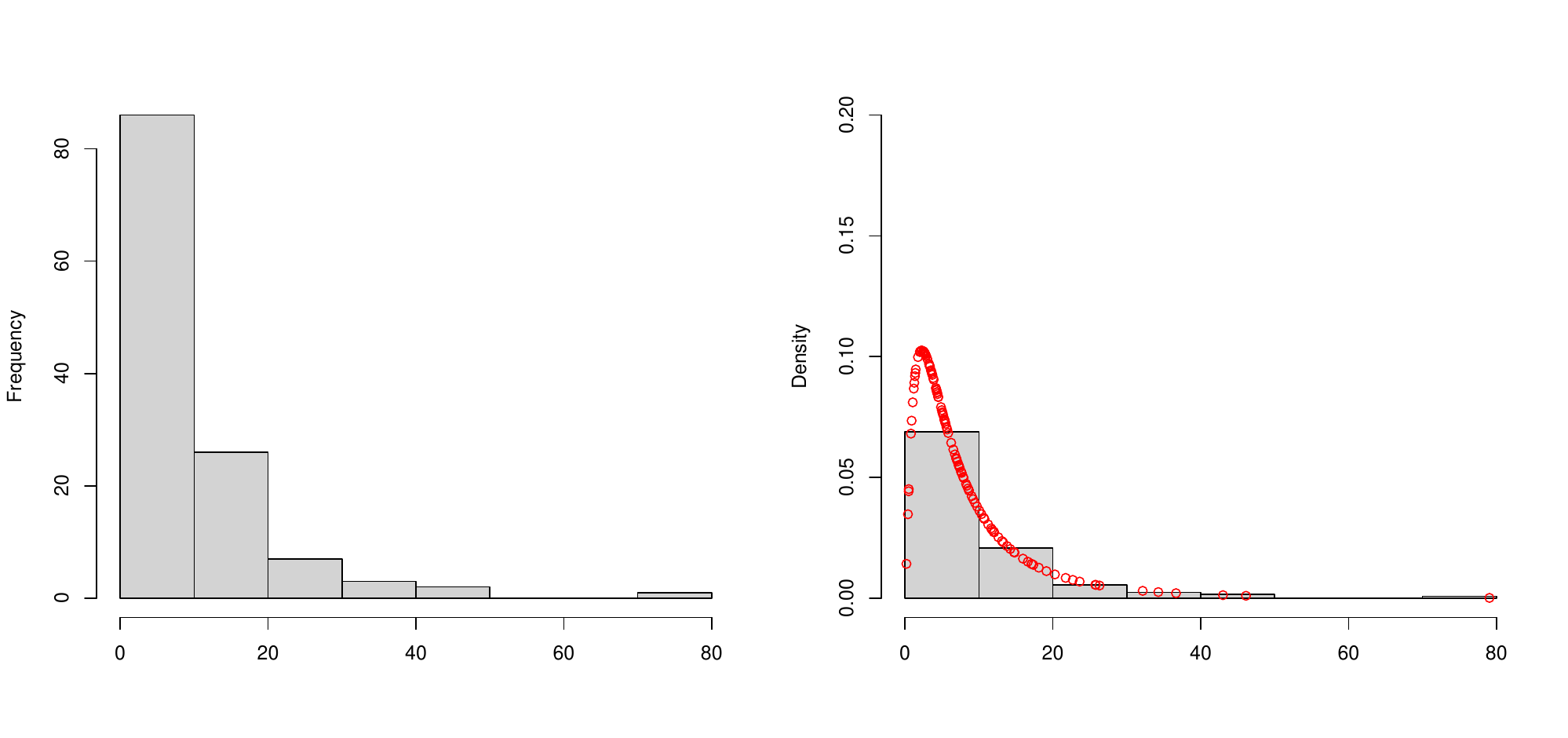}
\caption{Histogram and density plot of remission times of 128 bladder cancer patients} \label{figure2}
\end{figure}

For GW model, the MLEs of the shape parameters for the data set are found to be $\hat{\theta}=0.470$ and $\hat{\alpha}=6.941$ respectively with Kolmogorov–Smirnov test (K-S) statistic value $D = 0.043,$ and p-value, $p = 0.971.$ The histogram of the data set and the fitted density are provided in Figure \ref{figure2}. The fit results confirm that the GW distribution provides an excellent fit to the data set.\\
\begin{figure}[ht]
\centering
\includegraphics[height=5.5 cm,width=15 cm]{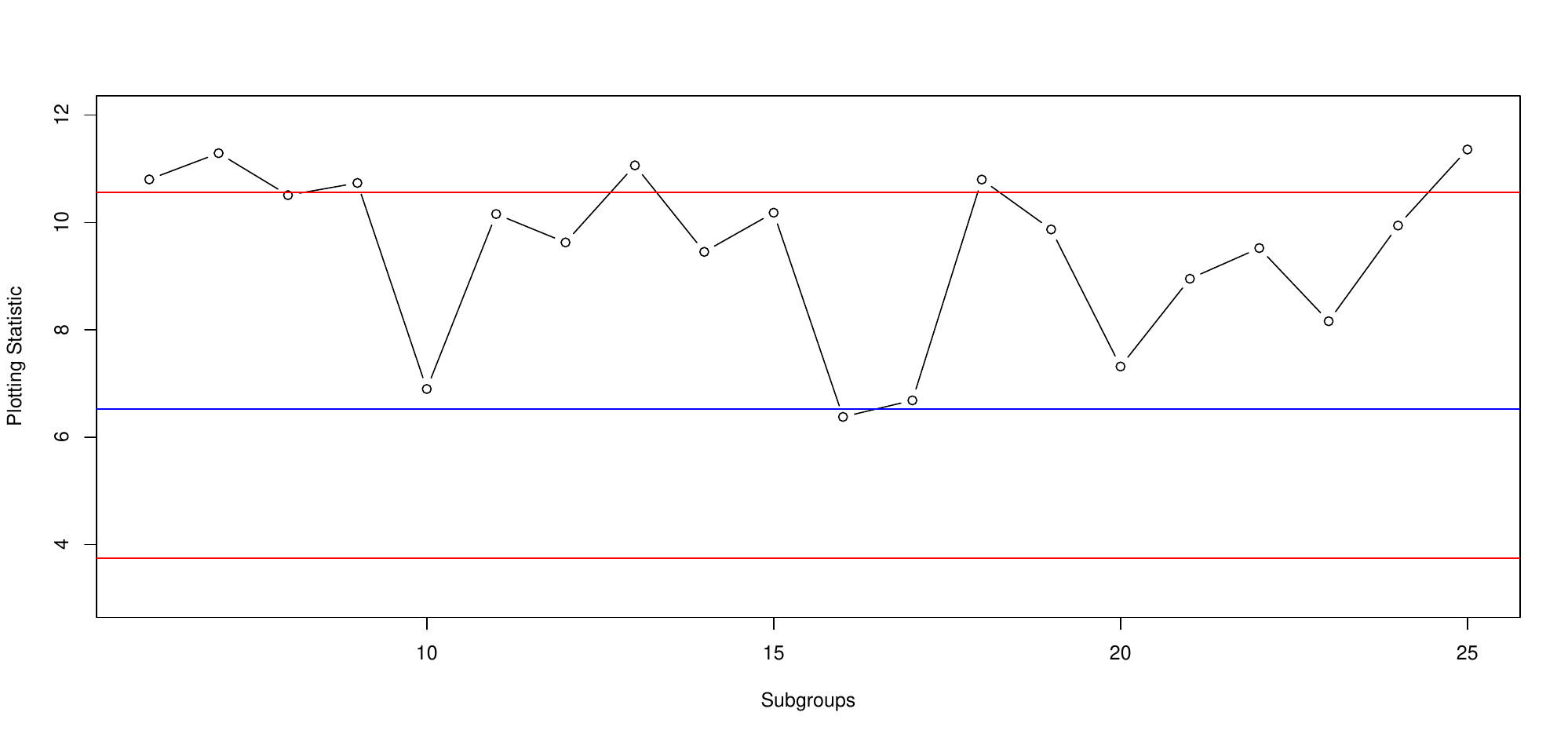}
\caption {BHC scheme for $90^{th}$ percentile of remission time data with $\Delta\alpha=0$, $\Delta\theta=-0.15$, $UCL=10.564$, $CL=6.524$, $LCL=3.742$} \label{figure3}
\end{figure}

\begin{figure}[ht]
\centering
\includegraphics[height=5.5 cm,width=15 cm]{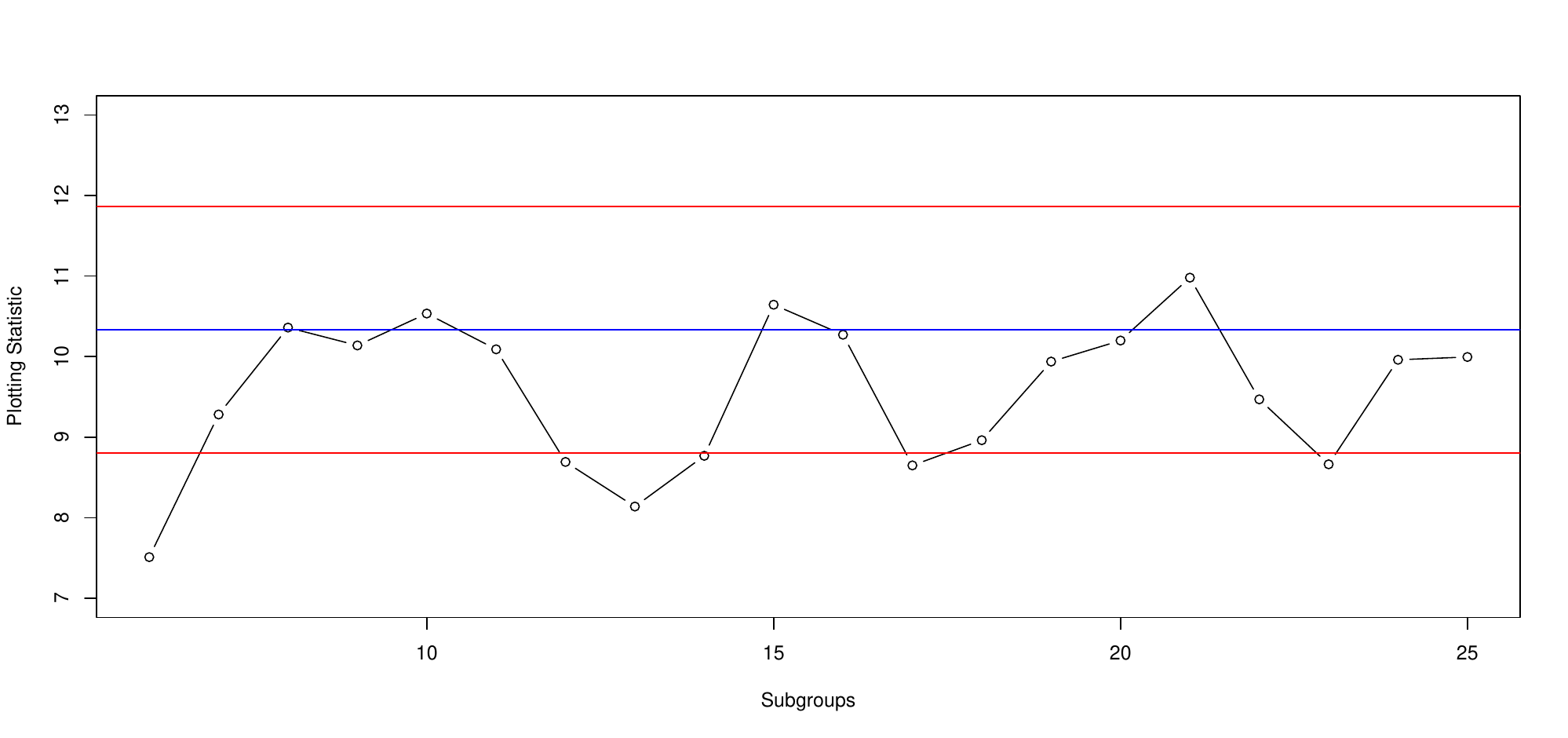}
\caption{SHC scheme for $90^{th}$ percentile of remission time data with $\Delta\alpha=0$, $\Delta\theta=-0.15$, $UCL_{SH}=11.864$, $CL_{SH}=10.333$, $LCL_{SH}=8.802$} \label{figure4}
\end{figure}
\hspace*{0.2in} For the purpose of monitoring cancer patients, it may be worthwhile to investigate extreme quantiles of the remission times over the average. An upward shift in the upper quantile of the remission times may indicate deterioration in the health status and may require monitoring. In view of this objective, the BHC scheme is proposed to monitor $90^{th}$ percentile of the remission times. The complete data is censored either at the remission time of the first $60\%$ of the patients $(r = 75)$ or at the remission time of $7.6$ $(x_0 = 7.6)$ months, whichever occurs earlier. The censoring time $x_0 = 7.6$ is chosen near to $60^{th}$ percentile. The complete set of $125$ observations is split into five $(k = 5)$ reference samples of size $m= 25$ each. The MLEs of $\theta$ and $\alpha$ under the stated hybrid censoring scheme are obtained as $\hat{\theta}$= 0.632 and $\hat{\alpha}= 8.946$ respectively. Using these MLEs, $B =5,000$ bootstrap samples of size $m= 25$ each are drawn with $r= 15$ $(60\%$ of the subgroup size) and $x_0 =7.6$.  Following the steps $4$-$7$ of subsection $3.1,$  and using $\nu=0.0027$ as FAR, the control limits of the BHC scheme for the $90^{th}$ quantile are obtained as $UCL= 10.564$,  $CL = 6.524$ and  $LCL = 3.742.$ Following the method described in subsection $3.2,$ the control limits of the SHC scheme is found to be $UCL_{SH}=11.864,$ $CL_{SH}=10.333$ and $LCL_{SH}=8.802.$ It is observed that the BHC scheme provides asymmetric control limits from the respective CL. Moreover, the SHC scheme has narrower interval than the BHC scheme. Twenty subgroups of size $m=25$ each are generated from the OOC process under similar hybrid censoring plan having shape parameters $\theta = 0.537$ ($15\%$ decrease in $\theta$) and in-control parameter $\alpha=8.946.$ \\
\hspace*{0.2in} The OOC performance of the BHC and SHC schemes is found to be quite competitive as provided in Figure $\ref{figure3}$ and Figure $\ref{figure4}.$ The BHC scheme produces six OOC signals with the first two signals produced at test sample $1$ and $2$, while SHC scheme generates first OOC signal at test sample $1$ out of six under the same shift in the parameters. Both the schemes are proven to be effective in detecting OOC signals not only in the magnitude but also in the speed. \\
\begin{figure}[ht]
\centering
\includegraphics[height=5.5 cm,width=15 cm]{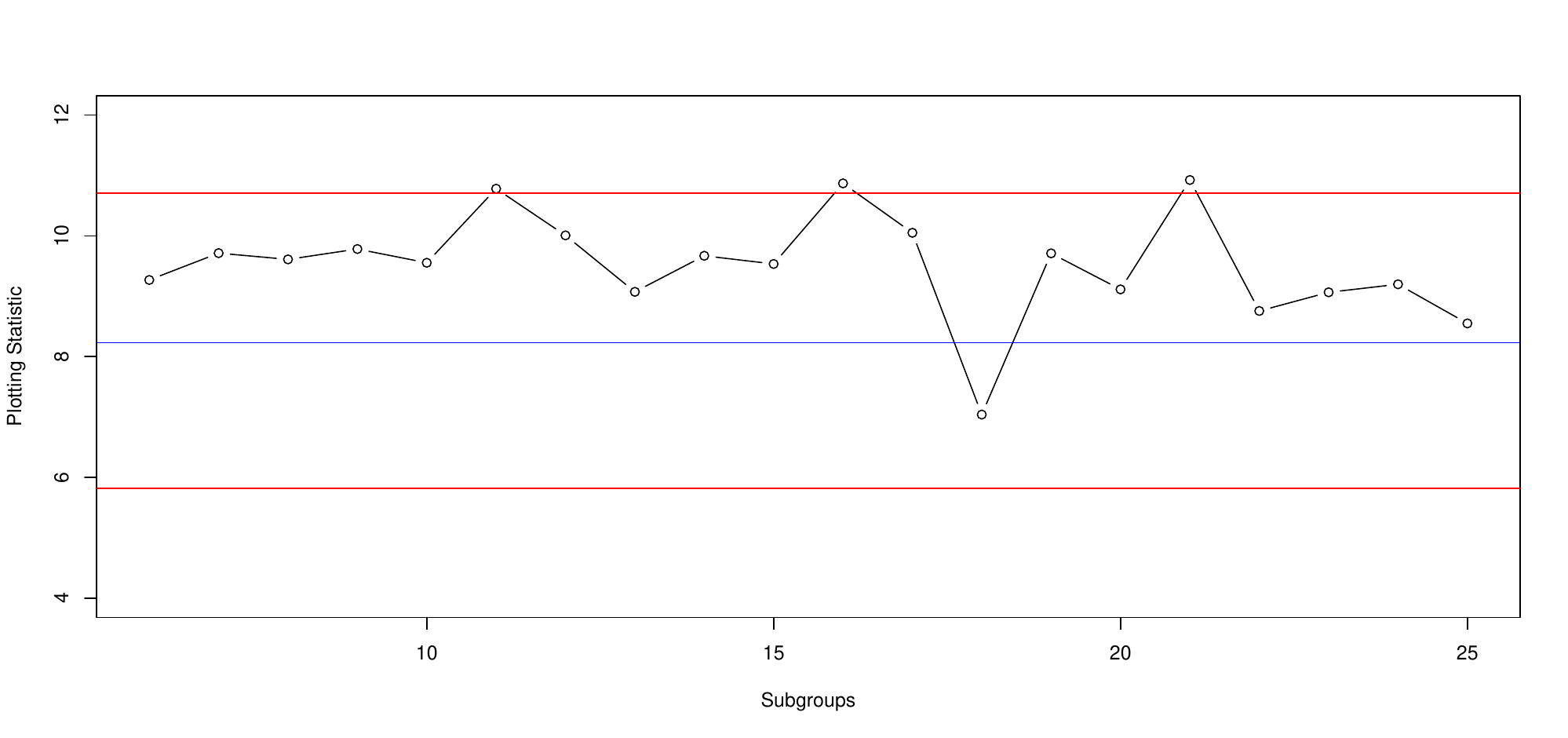}
\caption{ BT$^I$C scheme for $90^{th}$ percentile of remission time data with $\Delta\alpha=0$, $\Delta\theta=-0.15$, $UCL=10.708$, $CL=8.232$, $LCL=5.819$} \label{figure5}
\end{figure}

\begin{figure}[ht]
\centering
\includegraphics[height=5.5 cm,width=15 cm]{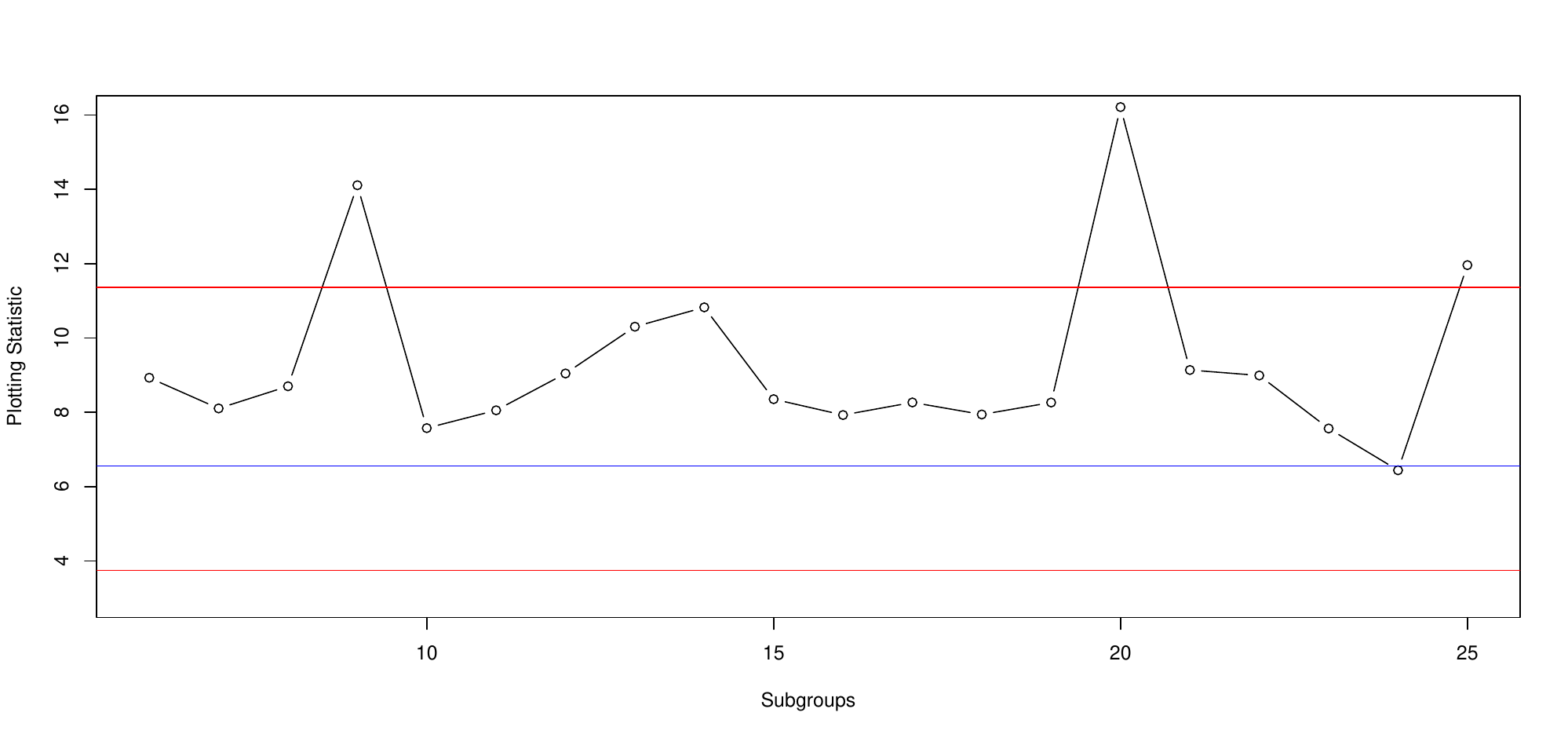}
\caption{ BT$^{II}$C scheme for $90^{th}$ percentile of remission time data with $\Delta\alpha=0$, $\Delta\theta=-0.15$, $UCL=11.361$, $CL=6.555$, $LCL=3.751$} \label{figure6}
\end{figure}
\hspace*{0.2in} Next, the BHC scheme is compared with the bootstrap schemes for type-I (denoted as $BT^IC$) and type-II (denoted as BT$^{II}$C) censored data which are obtained as the special cases of hybrid censored data for $r=n$ and $T=x_{n:n}$ respectively. The MLEs of $\theta$ and $\alpha$ under type-I censoring with $x_0 = 7.6$ are obtained as $\hat{\theta} = 0.632$ and $\hat{\alpha} = 8.946$ respectively, while the same under type-II censoring with $r = 75$ ($60\%$ of the total data set) are found to be $\hat{\theta} = 0.630$ and $\hat{\alpha} = 8.909$ respectively. The control limits of BT$^I$C scheme for the $90^{th}$ percentile are obtained as $UCL = 10.708$, $CL = 8.232$ and $LCL = 5.819$ while the same for the BT$^{II}$C scheme are calculated as $UCL = 11.361$, $CL = 6.555$, and $LCL = 3.751.$ Both the schemes provide asymmetric control limits with respect to CL with the BT$^{I}$C scheme having narrower control limits than the BT$^{II}$C scheme. After the first five IC subgroups, twenty subgroups of size $m = 25$ each are generated from the OOC process with the same shift in the parameters as of BHC scheme. Figure $\ref{figure5}$ and Figure $\ref{figure6}$ provide the OOC performance of the control schemes for the $90^{th}$ quantile of BT$^I$C and B$T^{II}C$ respectively. Figure $\ref{figure5}$ shows that the B$T^{I}$CC scheme is able to generate three OOC signals falling above the UCL with the first being produced at test sample $6$. The B$T^{II}$C scheme scheme as shown in Figure $\ref{figure6}$ produces three OOC signals just above the UCL with test sample $4$ providing the first signal. It is evident from the data analysis that the bootstrap and Shewhart type schemes under hybrid censored scheme perform better than the conventional type I and type II censored schemes in terms of both frequency and speed of detection of OOC signals.

\setcounter{section}{5} 
\section{Concluding Remark}
\setcounter{equation}{0}
\hspace*{0.2in} In the present work, bootstrap and Shewhart type control schemes are proposed for monitoring quantiles of generalized Weibull distribution under hybrid censoring. Maximum likelihood estimators of the parameters of the underlying distribution are derived for hybrid censored data using EM algorithm and their asymptotic properties are discussed to develop the Shewhart type scheme. Control schemes for the quantiles of Weibull, generalized exponential, Rayleigh, and Burr type $X$ distributions for type I, type II and hybrid censoring can be developed as the special cases of the proposed schemes. In this sense, the present work is an attempt to apply a new censoring scheme in the process control and generalizes available monitoring schemes. An extensive simulation study is conducted to evaluate the in-control and out-of-control performance of the schemes proposed. The bootstrap and Shewhart type schemes are found to be highly effective in the detection of out-of-control signals in terms of both magnitude and speed as demonstrated by a skewed data set from healthcare.

\begin{center}
\tiny{
\begin{longtable}{|c|c|c|c|c|c|c|}
\caption{Control limits, $ARL_0$ and $SDRL_0$ for $\theta=0.51, \alpha=11.1$}\label{tab1}\\
\hline
\multicolumn{7}{|c|}{$m=25,$ $x_0=55,$ $r=15$}\\ 
\hline
\endfirsthead
\multicolumn{7}{l}%
{\tablename\ \thetable\ -- \textit{Continued from previous page}} \\
\hline
 $\nu$ & $\theta$ & $\alpha$ & LCL & UCL & $ARL_0$ & $SDRL_0$\\  
\hline
\endhead
\hline \multicolumn{7}{r}{\textit{Continued on next page}} \\
\endfoot
\hline
\endlastfoot
$\nu$ & $\theta$ & $\alpha$ & LCL & UCL & $ARL_0$ & $SDRL_0$\\
\hline
\multicolumn{7}{|c|}{$p=0.1$}\\
\hline
0.005	& 0.654	& 17.297	& 2.241	& 4.830	& 195.799	& 193.848\\
\hline
0.0027	& 16.726	& 0.649	& 2.133  & 4.935	& 363.124	& 357.056\\
\hline
0.002	& 16.497	& 0.653	& 0.653	& 4.908	& 497.775	& 496.052\\
\hline
\multicolumn{7}{|c|}{$p=0.5$}\\
\hline																
0.005	& 17.171	& 0.656	& 3.710		& 7.035	& 202.123	& 202.868	\\
\hline
0.0027 & 17.758 & 0.656	& 3.662	& 7.245	& 368.720	& 367.574	\\
\hline
0.002	&	17.374	&	0.658	&	3.601	&	7.281	& 503.662	& 504.234\\	
\hline
\multicolumn{7}{|c|}{$p=0.9$}\\
\hline
0.005	& 17.320	& 0.658	& 5.448	& 12.558	& 203.946	& 202.924\\
\hline
0.0027	& 17.029	& 0.662	& 5.190	& 12.731 & 378.223 & 377.664\\
\hline
0.002	& 16.670	& 0.651	& 5.203	& 13.534	& 512.959	& 506.114\\
\hline

\multicolumn{7}{|c|}{$m=25,$ $x_0=55$, $r=20$}\\
\hline
\multicolumn{7}{|c|}{$p=0.1$}\\
\hline
0.005	& 13.686	& 0.577& 	1.9486 & 	5.158 & 205.481	& 201.574\\
\hline
0.0027 &	14.101 & 	0.577	& 1.929	& 5.354	& 374.595	& 373.981\\
\hline
0.002	& 13.659	& 0.574	& 1.843 & 	5.381	& 505.957	& 504.177\\
\hline

\multicolumn{7}{|c|}{$p=0.5$}\\
\hline
0.005	& 13.281	& 0.573	& 4.224	& 8.924	& 195.560	& 195.365\\
\hline
0.0027	& 13.563	& 0.571	 & 4.212	& 9.339	& 372.126	& 367.878\\
\hline
0.002	& 13.692	& 0.573	& 4.165	& 9.425	& 487.093	& 486.439\\
\hline
\multicolumn{7}{|c|}{$p=0.9$}\\
\hline
0.005	& 13.522	& 0.570	& 7.752	& 20.7741	& 201.5118 & 200.054\\
\hline
0.0027	& 13.586	& 0.572	& 7.445	& 21.193 & 368.971	& 362.621\\
\hline
0.002	& 13.560	& 0.571	& 7.354	& 21.687	& 497.131	& 485.442\\
\hline

\multicolumn{7}{|c|}{$m=40,$ $x_0=55,$ $r=30$}\\
\hline
\multicolumn{7}{|c|}{$p=0.1$}\\
\hline
0.005	& 14.095	& 0.590	& 2.182	& 4.552	& 198.473	& 196.097\\
\hline
0.0027 & 	13.993	& 0.591	& 2.106	& 4.626	& 371.156	& 368.505\\
\hline
0.002	& 14.522 & 0.589	& 2.162 & 4.822	& 499.847	& 488.881\\
\hline

\hline
\multicolumn{7}{|c|}{$p=0.5$}\\
\hline
0.005	& 14.141	& 0.591	& 4.442	& 7.837	& 201.168	& 199.661\\
\hline
0.0027	& 14.338	& 0.591	 & 4.395	& 8.037	& 369.931	& 364.105\\
\hline
0.002	& 14.139	& 0.589	& 4.337	& 8.121	& 496.312	& 496.331\\
\hline

\hline
\multicolumn{7}{|c|}{$p=0.9$}\\
\hline
0.005	& 14.263	& 0.591	& 7.804	& 16.430	& 198.135	& 196.715\\
\hline
0.0027	& 14.527	& 0.593	& 7.637	& 16.820	& 363.512	& 360.149\\
\hline
0.002	& 14.263	& 0.590	& 7.577	& 17.231	& 494.709	& 489.603\\
\hline
\multicolumn{7}{|c|}{$m=40,$ $x_0=55,$ $r=35$}\\
\hline
\multicolumn{7}{|c|}{$p=0.1$}\\
\hline
0.005	& 12.638	& 0.550	& 1.994	& 4.603	& 205.850	& 204.971\\
\hline
0.0027	& 12.441	& 0.544	& 1.916	& 4.758	& 366.314	& 364.314\\
\hline
0.002	& 12.491	& 0.547	& 1.888	& 4.784	& 485.950	& 485.090\\
\hline
\multicolumn{7}{|c|}{$p=0.5$}\\
\hline
0.005	& 12.638	& 0.550	& 1.994	& 4.603	& 205.850	& 204.971\\
\hline
0.0027	& 12.441	& 0.544	& 1.916	& 4.758	& 366.314	& 364.314\\
\hline
0.002	& 12.491	& 0.547	& 1.888	& 4.784	& 485.950	& 485.090\\
\hline

\hline
\multicolumn{7}{|c|}{$p=0.9$}\\
\hline
0.005	& 12.518	& 0.549	& 9.802	& 22.490	& 198.070	& 197.586\\
\hline
0.0027	& 12.849	& 0.550	& 9.665	& 23.282	& 372.609	& 372.181\\
\hline
0.002	& 12.731	& 0.549	& 9.508	& 23.528	& 498.132	& 494.361 \\
\hline
\multicolumn{7}{|c|}{$m=25,$ $x_0=70,$ $r=15$}\\
\hline
\multicolumn{7}{|c|}{$p=0.1$}\\
\hline
0.005	& 16.814	& 0.654	& 2.189	& 4.757	& 198.900	& 196.694\\
\hline
0.0027	& 17.005	& 0.656	& 2.144	& 4.885 & 366.685	& 360.209\\
\hline
0.002	& 17.002	& 0.655	& 2.120	& 4.962	& 482.258	& 483.258\\
\hline
\multicolumn{7}{|c|}{$p=0.5$}\\
\hline
0.005	& 16.906	& 0.653	& 3.694	& 7.050	& 200.783	& 200.191\\
\hline
0.0027	& 17.414	& 0.656	& 3.659	& 7.249	& 374.606	& 365.391\\
\hline
0.002	& 17.007	& 0.651	& 5.269	& 13.709	& 538.988	539.441\\
\hline

\hline
\multicolumn{7}{|c|}{$p=0.9$}\\
\hline
0.005	& 16.829	& 0.656	& 5.394	& 12.543	& 195.944	& 194.987\\
\hline
0.0027	& 16.889	& 0.654	& 5.278	& 13.135	& 381.341	& 381.013\\
\hline  
0.002	& 16.632	& 0.654	& 5.159	& 13.305	& 511.462	& 509.388\\
\hline
\multicolumn{7}{|c|}{$m=25,$ $x_0=70,$ $r=20$}\\
\hline
\multicolumn{7}{|c|}{$p=0.1$}\\
\hline
0.005	& 13.657	& 0.573	& 1.944	& 5.148	 & 196.448	& 196.506\\
\hline
0.0027	& 13.526	& 0.571	& 1.866	& 5.329	& 372.386 & 368.807\\
\hline
0.002	& 13.642	& 0.572	& 1.846	& 5.408	& 494.614	& 492.127\\
\hline
\multicolumn{7}{|c|}{$p=0.5$}\\
\hline
0.005	& 13.443	& 0.571	& 4.284	& 9.044	& 197.523	& 192.749\\
\hline
0.0027	& 13.538	& 0.573	& 4.177	& 9.243	& 371.082	& 362.144\\
\hline
0.002	& 13.825	& 0.573	& 4.196	& 9.471	& 494.345	& 493.169\\
\hline

\multicolumn{7}{|c|}{$p=0.9$}\\
\hline
0.005	& 13.103	& 0.574	& 7.495	& 20.089	& 204.431	& 204.812\\
\hline
0.0027	& 13.248	& 0.573	& 7.339	& 21.021	& 381.413	& 381.971\\
\hline
0.002	& 13.645	& 0.575	& 7.289	& 21.314	& 508.737	& 506.744\\
\hline

\multicolumn{7}{|c|}{$m=40,$ $x_0=70,$ $r=30$}\\
\hline
\multicolumn{7}{|c|}{p=$0.1$}\\
\hline
0.005	& 14.053	& 0.590	& 2.176	& 4.540	& 204.173	& 203.714\\
\hline
0.0027	& 14.121	& 0.588	& 2.134	& 4.695	& 371.266	& 368.033\\
\hline
0.002	& 14.519	& 0.593	& 2.150	& 4.772	& 502.184	& 489.232\\
\hline
\multicolumn{7}{|c|}{$p=0.5$}\\
\hline
0.005	& 14.264	& 0.591	& 4.480	& 7.899	& 199.321	198.939\\
\hline
0.0027 &	14.273 & 	0.590	& 4.398	& 8.070	& 371.020	& 370.319\\
\hline
0.002	& 14.519	& 0.593	& 2.150	& 4.772	& 502.184	& 489.232\\
\hline

\hline
\multicolumn{7}{|c|}{$p=0.9$}\\
\hline
0.005	& 14.384	& 0.591	& 7.847	& 16.498	& 194.875	& 190.947\\
\hline
0.0027	& 14.405	& 0.593	& 7.608	& 16.792	& 370.997	& 371.897\\
\hline
0.002	& 13.972	& 0.588	& 7.530	& 17.240	& 506.998	& 500.142\\
\hline
\multicolumn{7}{|c|}{$m=40,$ $x_0=70,$ $r=35$}\\
\hline
\multicolumn{7}{|c|}{$p=0.1$}\\
\hline
0.005	& 12.473	& 0.549	& 1.966	& 4.566	& 200.660	& 201.841\\
\hline
0.0027	& 12.631	& 0.548	& 1.939	& 4.762	& 370.184	& 370.390\\
\hline
0.002	& 12.620	& 0.549	& 1.906	& 4.801	& 514.149	& 509.982\\
\hline

\hline
\multicolumn{7}{|c|}{$p=0.5$}\\
\hline
0.005	& 12.768	& 0.549	& 4.953	& 9.173	& 196.905	& 197.722\\
\hline
0.0027	& 12.551	& 0.548	& 4.800	& 9.323	& 389.001	& 389.011\\
\hline
0.002	& 12.368	& 0.546	& 4.735	& 9.432	& 516.985	& 507.353\\
\hline

\hline
\multicolumn{7}{|c|}{$p=0.9$}\\
\hline
0.005	& 12.737	& 0.551	& 9.826	& 22.447	& 203.651	& 203.249\\
\hline
0.002	& 12.500	& 0.051	& 9.332	& 23.121	& 500.438	& 492.649\\
\hline
0.0027	& 12.537	& 0.544	& 9.751	& 23.741	& 364.350	& 363.350\\
\hline

\end{longtable}}
\end{center}

\subsection{OOC Chart Performance}

\begin{center}
\tiny{
\begin{longtable}{|c|c|c|c|}
\caption{OOC performance for $m=25,$ $x_0=55,$ $r=15,$ $\nu=0.0027$}\label{tab2}\\
\hline
$\Delta\theta$ & $p=0.1$ & $p=0.5$ & $p=0.9$\\
 \hline
\endfirsthead
\multicolumn{4}{l}
{\tablename\ \thetable\ -- \textit{Continued from previous page}} \\
\hline
$\Delta\theta$ & $p=0.1$ & $p=0.5$ & $0.9$\\
\hline
\endhead
\hline \multicolumn{4}{r}
{\textit{Continued on next page}} \\
\endfoot
\hline
\endlastfoot
\multicolumn{4}{|c|}{$\Delta\theta=-0.2$}\\
\hline
 & ARL(SDRL) & ARL(SDRL) & ARL(SDRL)\\
\hline
\multicolumn{4}{|c|}{$\Delta\alpha=-0.2$}\\
\hline
 -0.1 & 45.007(44.051) & 53.358(52.970) & 45.018 (45.002)\\ 
\hline
        -0.08 & 60.455(60.259) & 98.638(98.141) & 70.915(69.765)\\
				\hline
        -0.06 & 48.451(48.958) & 119.533(122.005) & 129.184(129.734)\\
				\hline
        -0.04 & 40.559(41.023) & 92.247(94.301) & 90.542(88.775)\\ 
				\hline
        -0.02 & 31.826(31.964) & 57.092(57.443) & 60.620(58.292)\\
				\hline
        0 & 26.418(26.796) & 34.066(33.912) & 73.838(72.0157)\\
				\hline
        0.02 & 20.085(20.097) & 21.569(22.091) & 44.808(45.382)\\
				\hline
        0.04 & 16.897(17.504) & 14.051(14.717) & 28.062(28.613)\\
				\hline
        0.06 & 13.682(14.254) & 8.949(9.483) & 17.199(17.468)\\
				\hline
        0.08 & 11.057(11.355) & 6.046(6.372) & 11.195(11.704)\\
				\hline
        0.1 & 9.231(9.701) & 4.186(4.676) & 7.348(7.954)\\
				\hline
        0.2 & 3.796(4.187) & 0.695(1.113) & 1.299(1.715)\\
				\hline
    
\multicolumn{4}{|c|}{$\Delta\alpha=-0.1$}\\
\hline
-0.2 & 6.807(7.294) & 0.994(1.362) & 1.028(1.426)\\
 \hline
        -0.08 & 41.689(41.114) & 42.542(43.483) & 38.156(37.897)\\
				\hline
        -0.06 & 151.262(153.159) & 95.911(94.190) & 85.436(85.465)\\ 
				\hline
        -0.04 & 163.508(160.806) & 186.547(186.592) & 174.309(177.353)\\        \hline
        -0.02 & 146.842(145.108) & 221.416(227.001) & 251.285(252.213)\\         \hline
        0 & 123.619(123.004) & 159.149(162.714) & 206.030(202.432)\\
				\hline
        0.02 & 93.748(94.048) & 94.357(93.061) & 120.096(120.148)\\
				\hline
        0.04 & 73.597(72.304) & 52.877(52.824) & 70.639(70.829) \\
				\hline
        0.06 & 56.244(5.183) & 31.404(31.697) & 42.367(42.986)\\ 
				\hline
        0.08 & 44.141(45.155) & 19.048(19.239) & 24.980(24.992)\\ 
				\hline
        0.1 & 34.336(35.579) & 12.224(12.424) & 15.554(15.974)\\ 
				\hline
        0.2 & 11.144(11.617) & 1.654(2.057) & 2.188(2.618)\\ 
				\hline
\multicolumn{4}{|c|}{$\Delta\alpha=-0.08$}\\
\hline
 -0.2 & 5.509(5.932) & 0.841(1.255) & 0.922(1.378)\\
 \hline
        -0.1 & 61.225(63.313) & 16.367(16.915) & 17.164(17.538)\\
				\hline
        -0.06 & 149.067(148.897) & 79.386(79.962) & 78.768(78.826)\\
				\hline
        -0.04 & 193.525(194.116) & 166.597(166.451) & 163.288(163.295)\\         \hline
        -0.02 & 194.210(194.933) & 250.529(254.097) & 259.701(257.398)\\        \hline
        0 & 262.578(260.639) & 214.156(209.551) & 244.595(242.505)\\
				\hline
        0.02 & 129.807(130.35) & 127.616(128.310) & 151.585(152.147.)\\         \hline
        0.04 & 99.658(98.038) & 72.131(71.647) & 87.334(87.581)\\
				\hline
        0.06 & 78.633(76.914) & 41.372(42.693) & 49.756(49.955)\\ 
				\hline
        0.08 & 59.314(60.307) & 54.696(53.279) & 30.295(30.115)\\
				\hline
        0.1 & 45.884(46.498) & 24.547(24.938) & 18.543(18.314)\\
				\hline
        0.2 & 17.264(15.598) & 5.749(4.998) & 2.488(2.987) \\ 
\hline
\multicolumn{4}{|c|}{$\Delta\alpha=-0.06$}\\
\hline
-0.2 & 4.824(4.285) & 0.687(1.095) & 0.814(1.245) \\ 
\hline
        -0.1 & 51.602(50.919) & 13.746(13.102) & 15.777(15.286)\\ 
				\hline
        -0.08 & 84.755(84.84.311) & 28.285(28.029) & 32.458(33.080)\\           \hline
        -0.04 & 202.655(201.866) & 146.787(145.613) & 151.134(150.791)\\        \hline
        -0.02 & 230.428(230.555) & 260.708(260.157) & 271.198(271.640)\\        \hline
        0 & 222.379(222.649) & 264.672(260.246) & 281.860(281.371)\\
				\hline
        0.02 & 180.639(179.454) & 168.802(168.552) & 183.987(183.158)\\         \hline
        0.04 & 137.917(137.900) & 96.990(96.635) & 107.148(107.910) \\          \hline
        0.06 & 104.639(104.483) & 53.427(53.256) & 59.812(59.982)\\
				\hline
        0.08 & 79.406(78.630) & 31.784(31.899) & 34.763(34.216)\\
				\hline
        0.1 & 61.994(61.184) & 19.392(19.468) & 20.936(20.866)\\
				\hline
        0.2 & 18.363(19.216) & 2.322(2.727) & 2.666(2.182) \\ 
				\hline
\multicolumn{4}{|c|}{$\Delta\alpha=-0.04$}\\
\hline
 -0.2 & 4.065(3.557) & 0.610(0.989) & 0.728(1.118)\\
 \hline
        -0.1 & 41.807(41.476) & 11.429(10.828) & 13.798(13.284)\\
				\hline
        -0.08 & 70.874(70.248) & 23.329(22.872) & 30.985(29.179)\\
				\hline
        -0.06 & 126.192(123.973) & 52.968(52.572) & 61.721(61.430)\\
				\hline
        -0.02 & 260.638(259.460) & 256.948(248.055) & 266.231(264.644)\\ 
				\hline
        0 & 281.823(280.869) & 317.095(315.725) & 328.096(323.794)\\ 
				\hline
        0.02 & 234.529(231.143) & 226.644(224.415) & 232.573(231.817) \\
				\hline
        0.04 & 193.043(190.332) & 125.519(124.212) & 128.633(128.579) \\ 
				\hline
        0.06 & 146.169(145.112) & 70.654(70.774) & 74.231(73.564)\\
				\hline
        0.08 & 107.574(107.615) & 41.072(41.876) & 42.713(42.248)\\
				\hline
        0.1 & 82.353(82.906) & 24.035(24.681) & 25.762(25.391)\\
				\hline
        0.2 & 23.126(23.303) & 2.914(2.434) & 3.068(2.514)\\
				\hline
\multicolumn{4}{|c|}{$\Delta\alpha=-0.02$}\\
\hline
-0.2 & 3.426(3.018) & 0.493(0.8703) & 0.717(1.086)\\
 \hline
        -0.1 & 34.026(33.670) & 10.740(9.165) & 12.613(11.232)\\
				\hline
        -0.08 & 58.931(57.959) & 20.548(19.246) & 25.794(25.905)\\ 
				\hline
        -0.06 & 103.498(103.694) & 45.694(44.844) & 55.484(55.241)\\
				\hline
        -0.04 & 178.321(178.647) & 102.793(101.735) & 124.076(124.971) \\        \hline
        0 & 333.416(332.896) & 341.265(340.256) & 356.316(355.348)\\
				\hline
        0.02 & 318.634(317.966) & 302.843(299.998) & 269.311(269.109)\\         \hline
        0.04 & 251.523(251.228) & 180.511(178.885) & 153.729(153.713)\\         \hline
        0.06 & 202.627(198.567) & 95.521(94.209) & 87.754(86.729)\\
				\hline
        0.08 & 146.003(145.146) & 52.994(52.476) & 49.060(49.540)\\
				\hline
        0.1 & 110.978(109.163) & 31.535(31.120) & 28.336(28.504)\\
				\hline
        0.2 & 29.274(28.505) & 3.328(3.776) & 3.333(3.399)\\ 
				\hline
\multicolumn{4}{|c|}{$\Delta\alpha=0$}\\
\hline
 -0.2 & 2.885(2.386) & 0.451(0.794) & 0.629(0.992)\\
 \hline
        -0.1 & 28.007(27.504) & 8.340(8.943) & 11.461(11.882)\\
				\hline
        -0.08 & 47.862(47.822) & 16.666(15.125) & 22.637(23.244)\\
				\hline
        -0.06 & 88.716(87.918) & 35.783(34.424) & 50.173(50.112)\\ 
				\hline
        -0.04 & 153.560(152.617) & 82.972(82.833) & 112.589(111.375)\\
				\hline
        -0.02 & 265.492(265.60) & 200.721(200.534) & 242.867(241.948) \\
				\hline
        0.02 & 341.149(340.483) & 327.617(326.811) & 319.829(315.105)\\
				\hline
        0.04 & 311.370(310.541) & 235.674(234.842) & 189.611(188.972)\\
				\hline
        0.06 & 272.238(266.877) & 127.001(125.663) & 103.028(103.808)\\
				\hline
        0.08 & 204.426(202.599) & 67.903(67.006) & 57.805(57.163)\\
				\hline
        0.1 & 151.987(150.482) & 39.305(39.465) & 33.726(33.863)\\
				\hline
        0.2 & 38.425(37.923) & 3.848(3.281) & 3.742(3.233)\\ 
				\hline
\multicolumn{4}{|c|}{$\Delta\alpha=0.02$}\\
\hline
-0.2 & 2.562(2.944) & 0.389(0.726) & 0.256(0.298)\\ 
\hline
        -0.1 & 23.787(23.881) & 7.116(7.654) & 10.416(10.696)\\
				\hline
        -0.08 & 41.122(41.874) & 14.363(14.907) & 20.600(20.343)\\
				\hline
        -0.06 & 71.478(71.403) & 30.350(30.268) & 44.631(44.322) \\
				\hline
        -0.04 & 133.202(133.207) & 70.540(69.262) & 99.790(99.142)\\
				\hline
        -0.02 & 232.340(231.523) & 170.270(169.899) & 224.569(223.552) \\
				\hline
        0 & 331.104(329.045) & 360.629(359.673) & 320.567(320.594)\\
				\hline
        0.04 & 344.787(343.754) & 304.792(303.003) & 222.829(217.967) \\
				\hline
        0.06 & 262.193(262.554) & 167.178(165.024) & 121.801(120.376) \\
				\hline
        0.08 & 183.253(180.611) & 90.613(89.165) & 69.196(69.332)\\
				\hline
        0.1 & 94.828(93.884) & 51.299(51.426) & 39.692(39.668) \\
				\hline
        0.2 & 49.001(49.683) & 4.492(4.952) & 3.498(3.589)\\ 
				\hline
\multicolumn{4}{|c|}{$\Delta\alpha=0.04$}\\
\hline
-0.2 & 2.153(2.538) & 0.329(0.665) & 0.549(0.936) \\
 \hline
        -0.1 & 21.853(20.153) & 6.012(6.4330) & 9.158(9.815)\\
				\hline
        -0.08 & 33.416(33.114) & 11.882(10.490) & 18.551(18.865)\\
				\hline
        -0.06 & 60.336(60.164) & 25.605(25.897) & 40.366(39.203) \\
				\hline
        -0.04 & 107.493(107.517) & 58.616(58.658) & 90.615(89.874) \\
				\hline
        -0.02 & 203.097(203.147) & 143.764(142.550) & 166.770 (165.037) \\       \hline
        0 & 351.289(350.405) & 338.671(330.089) & 323.087(322.010)\\ 
				\hline
        0.02 & 255.426(254.088) & 209.072(208.816) & 208.562(207.905)\\
				\hline
        0.06 & 194.817(193.956) & 176.059(175.184) & 150.629(150.113) \\
				\hline
        0.08 & 124.271(123.755) & 118.062(117.499) & 121.691(120.678)\\
				\hline
        0.1 & 81.647(80.718) & 64.811(63.169) & 48.627(46.805)\\ 
				\hline
        0.2 & 15.311(15.302) & 5.358(5.797) & 4.481(4.016)\\
				\hline
\multicolumn{4}{|c|}{$\Delta\alpha=0.06$}\\
\hline
 -0.2 & 1.876(1.353) & 0.285(0.606) & 0.229 (0.302)\\
 \hline
        -0.1 & 16..650(15.373) & 5.278(4.696) & 0.466(0.841)\\
				\hline
        -0.08 & 27.687(27.257) & 10.876(10.707) & 16.567(15.124)\\
				\hline
				-0.04	& 202.655(201.866)	& 146.787(145.613)	& 151.134(150.791)\\
				\hline
        -0.02 & 168.657(165.508) & 119.796(118.189) & 192.024(190.692) \\ 
				\hline
        0 & 308.551(307.597) & 283.999(282.059) & 294.903(293.480)\\
				\hline
        0.02 & 226.005(225.723) & 203.421(202.229) & 205.859(204.526)\\
				\hline
        0.04 & 182.992 (181.071) & 157.045 (156.096) & 170.014 (170.132) \\        \hline
        0.08 & 112.285 (112.020) & 97.149(96.612) & 99.349(98.649) \\
				\hline
        0.1 & 59.772(58.770) & 42.752(42.964) & 53.963(52.442)\\
				\hline
        0.2 & 12.028(11.142) & 6.359(6.866) & 5.018(5.494) \\
				\hline
\multicolumn{4}{|c|}{$\Delta\alpha=0.08$}\\
\hline
-0.2 & 1.651(1.124) & 0.250(0.567) & 0.457(0.815)\\
 \hline
        -0.1 & 13.818(13.336) & 4.424(4.155) & 7.804(7.295)\\
				\hline
        -0.08 & 24.864(22.074) & 8.848(8.245) & 15.657(15.974)\\
				\hline
        -0.06 & 40.134(39.622) & 17.722(17.955) & 32.417(32.004) \\
				\hline
        -0.04 & 72.202(72.781) & 42.747(41.292) & 76.281(75.009) \\
				\hline
        -0.02 & 136.574(136.158) & 99.510(99.132) & 173.771(172.056)\\
				\hline
        0 & 258.524(258.719) & 237.069 (236.410) & 244.424(244.328) \\
				\hline
        0.02 & 177.599(176.926) & 153.554 (153.002) & 163.047 (162.951) \\      \hline
        0.04 & 159.020 (158.814) & 112.401(111.914) & 106.044(105.283)\\
				\hline
        0.06 & 109.610 (108.156) & 71.698(70.042) & 83.127(82.523)\\
				\hline
        0.1 & 84.186 (84.353) & 36.478(35.963) & 52.849(52.207)\\
				\hline
        0.2 & 21.028(20.142) & 7.342(7.925) & 5.639(6.227)\\
				\hline

\multicolumn{4}{|c|}{$\Delta\alpha=0.1$}\\
	\hline
-0.2 & 1.397(1.987) & 0.207(0.503) & 0.395(0.722)\\ 
\hline
        -0.1 & 11.861(12.464) & 3.972(4.268) & 8.174(7.369)\\
				\hline
        -0.08 & 19.458(20.368) & 7.441(8.040) & 14.381(14.593)\\ 
				\hline
        -0.06 & 33.489(34.851) & 15.527(16.137) & 28.703(28.243)\\ 
				\hline
        -0.04 & 60.803(62.208) & 34.364(35.394) & 66.320(67.538) \\
				\hline
        -0.02 & 114.580(117.593) & 82.047(81.981) & 157.763(162.637)\\
				\hline
        0 & 217.563(219.905) & 178.009(177.919) & 198.765(197.814)\\
				\hline
        0.02 & 198.257 (198.001) & 177.082(176.588) & 178.838 (178.737)\\ 
				\hline
        0.04 & 177.065 (176.798) & 134.046 (132.997) & 153.098 (152.901)\\      \hline
        0.06 & 109.552 (109.159) & 67.213 (67.225) & 90.708 (90.811)\\
				\hline
        0.08 & 51.010(50.690) & 23.115(23.150) & 39.265(38.758)\\
				\hline
        0.2 & 16.882(17.265) & 8.993(8.668) & 7.208(6.591)\\
				\hline
\hline
\multicolumn{4}{|c|}{$\Delta\alpha=0.2$}\\
\hline
-0.2 & 0.760(1.141) & 0.093(0.319) & 0.271(0.609)\\
 \hline
        -0.1 & 5.771(6.445) & 2.095(2.581) & 4.851(5.361)\\
				\hline
        -0.08 & 8.883(9.221) & 3.868(4.387) & 9.094(9.394)\\ 
				\hline
        -0.06 & 15.058(15.417) & 7.660(8.220) & 18.800(19.031)\\ 
				\hline
        -0.04 & 26.052(26.599) & 15.834(16.191) & 41.182(40.068)\\
				\hline
        -0.02 & 45.295(45.148) & 35.959(36.968) & 48.693(47.364)\\
				\hline
        0 & 85.360(84.699) & 87.780(87.016) & 89.142 (88.830)\\
				\hline
        0.02 & 64.338(64.372) & 53.786(53.409) & 68.901 (67.898)\\ 
				\hline
        0.04 & 57.879 (57.031) & 41.310 (41.100) & 49.487 (48.813)\\
				\hline
        0.06 & 39.089 (37.898) & 29.805 (29.384) & 12.520 (11.901)\\
				\hline
        0.08 & 22.118 (22.204) & 12.735 (12.201) & 8.550 (7.364)\\
				\hline
        0.1 & 9.108 (9.154) & 1.090 (1.841) & 3.401(3.832)\\
\end{longtable}}
\end{center}
\newpage
\begin{figure}[ht]
\centering
\begin{minipage}[b]{0.48\linewidth}
\includegraphics[height=4.2 cm, width=7.2 cm]{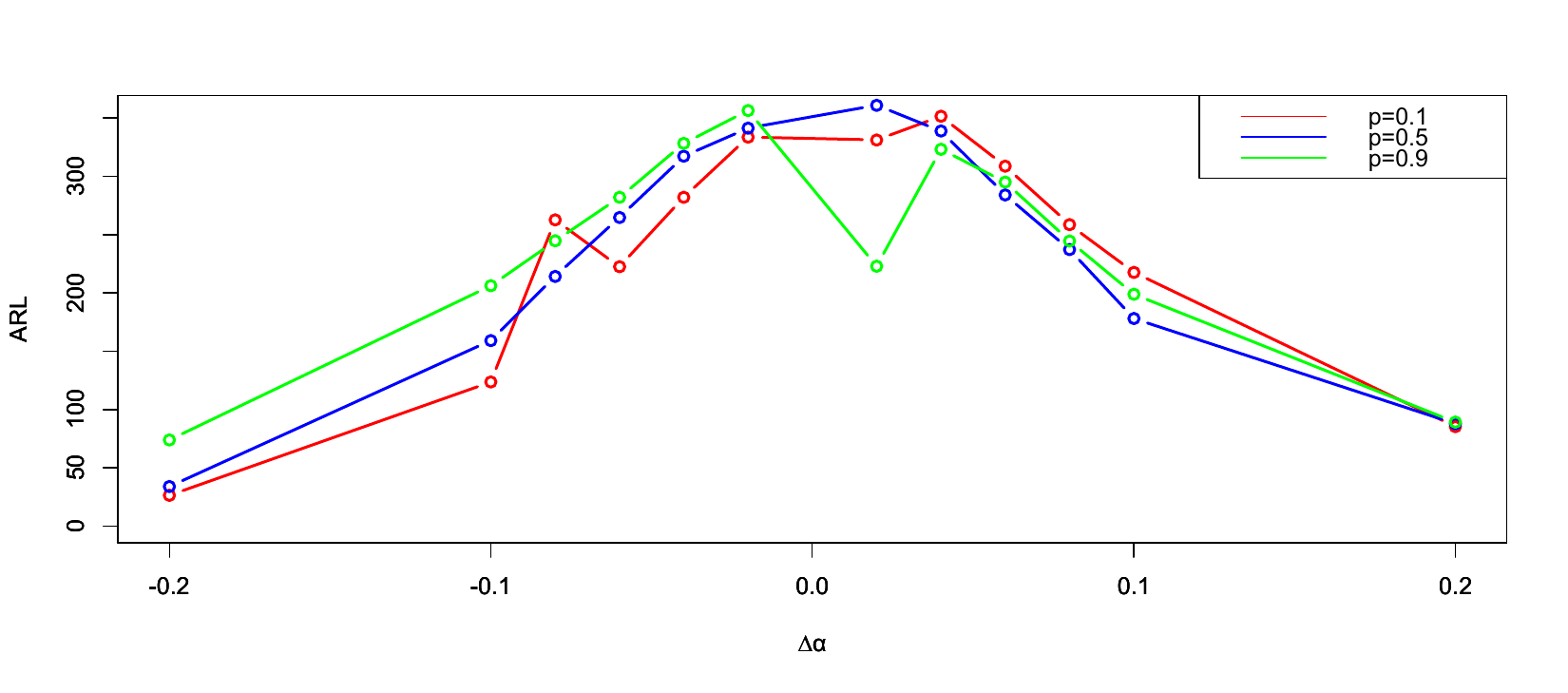}
\centering{$\left(a\right)$ $ARL_1$ for different choices of $\Delta\alpha$, when $\Delta\theta=0$ }
\end{minipage}
\quad
\begin{minipage}[b]{0.48\linewidth}
\includegraphics[height=4.2 cm, width=7.2 cm]{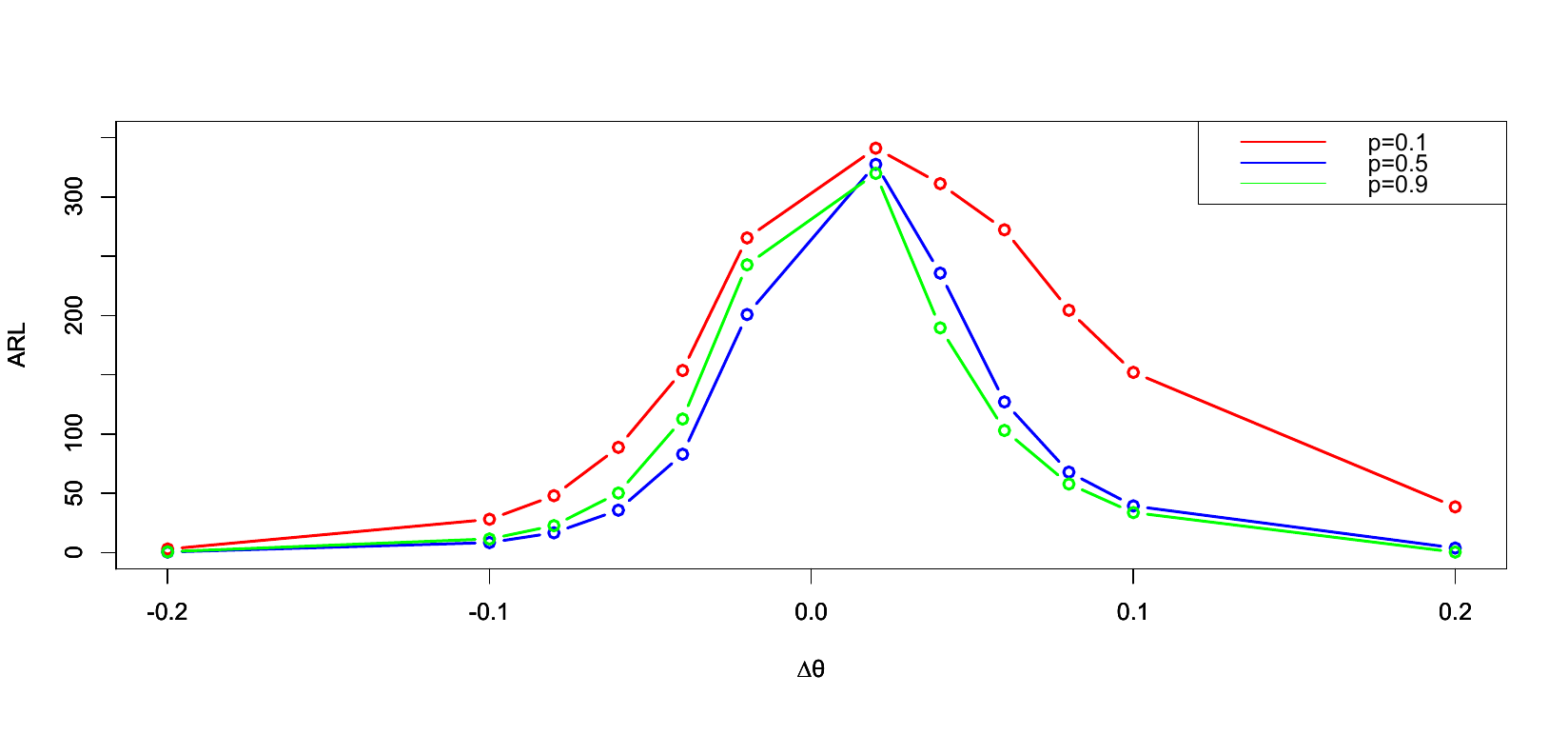}
\centering{$\left(b\right)$ $ARL_1$ for different choices of $\Delta\theta$, when $\Delta\alpha=0$}
\end{minipage}
\quad
\begin{minipage}[b]{0.48\linewidth}
\includegraphics[height=4.2 cm, width=7.2 cm]{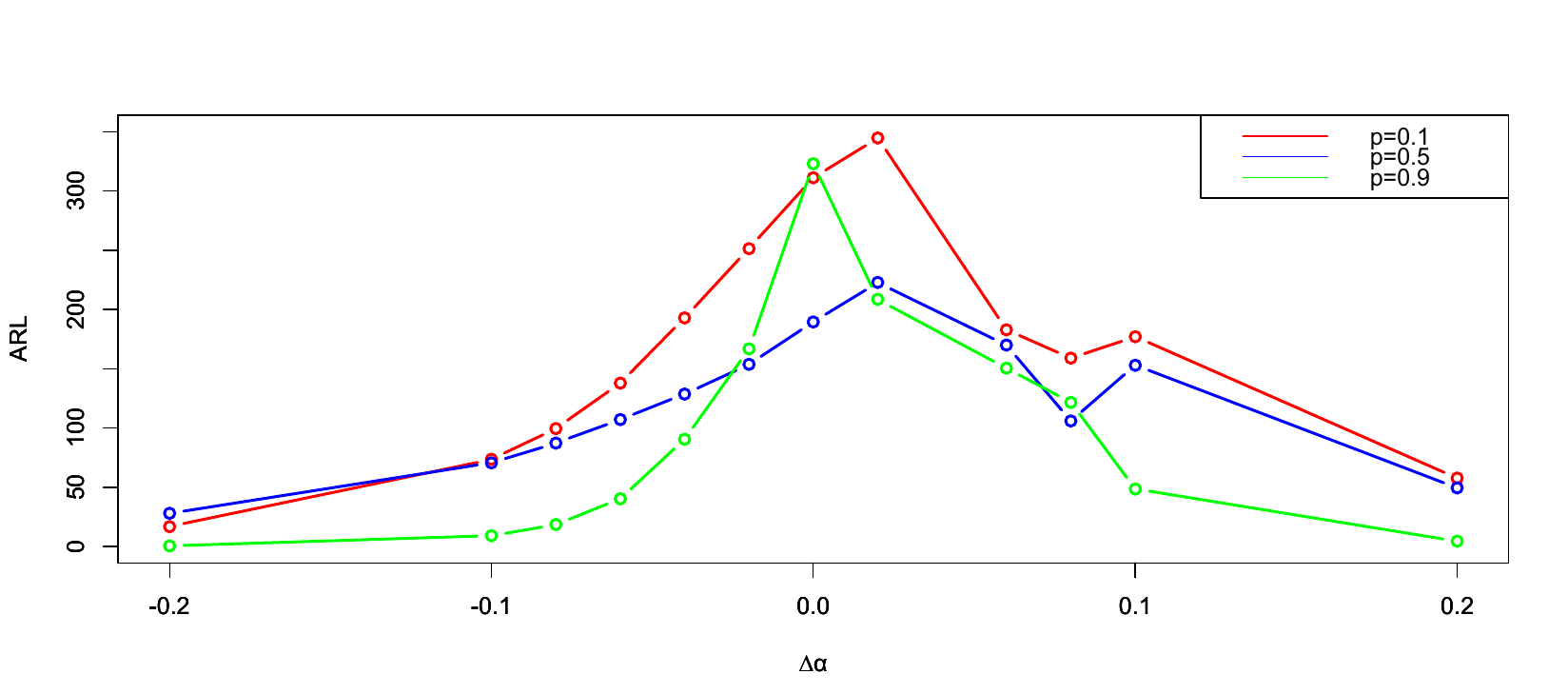}
\centering{$\left(c\right)$ $ARL_1$ for different choices of $\Delta\alpha$, when $\Delta\theta=0.04$}
\end{minipage}
\quad
\begin{minipage}[b]{0.48\linewidth}
\includegraphics[height=4.2 cm, width=7.2 cm]{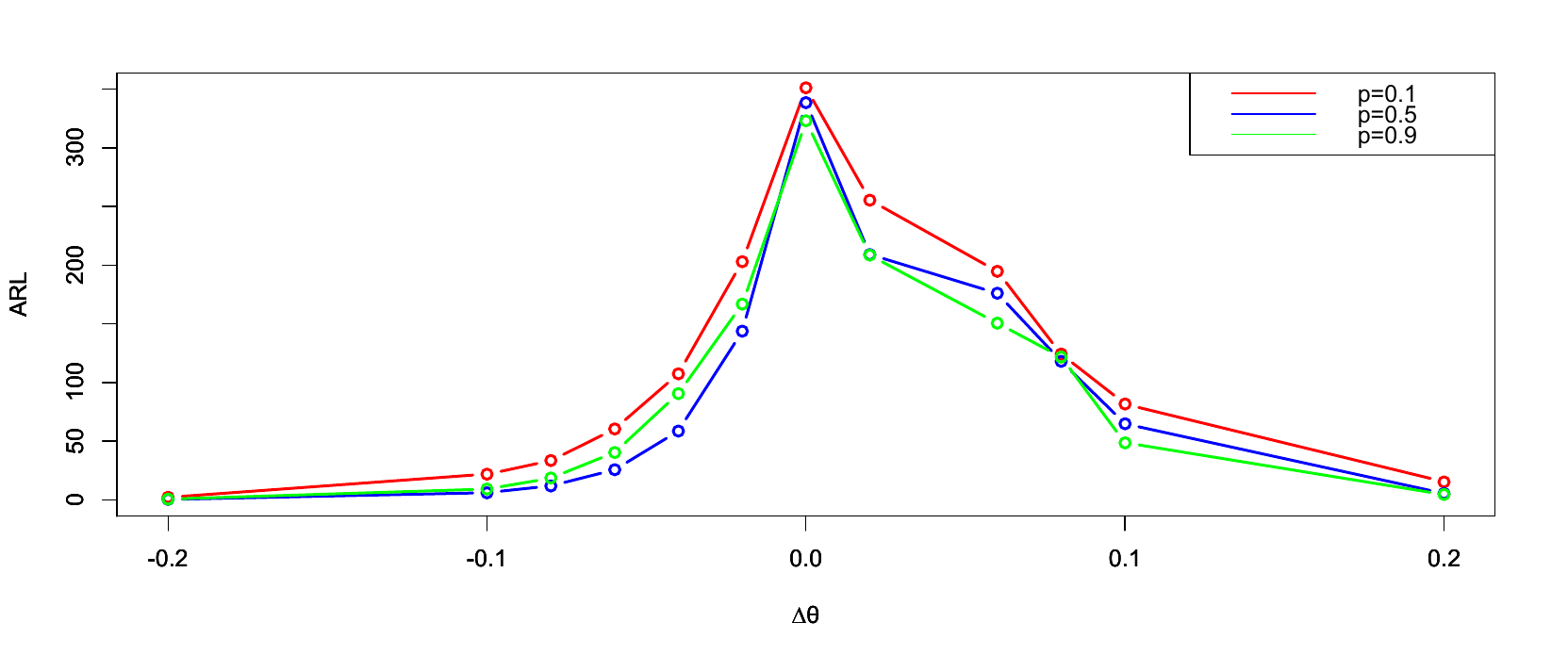}
\centering{$\left(d\right)$ $ARL_1$ for different choices of $\Delta\theta$, when $\Delta\alpha=0.04$}
\end{minipage}
\quad
\begin{minipage}[b]{0.48\linewidth}
\includegraphics[height=4.2 cm, width=7.2 cm]{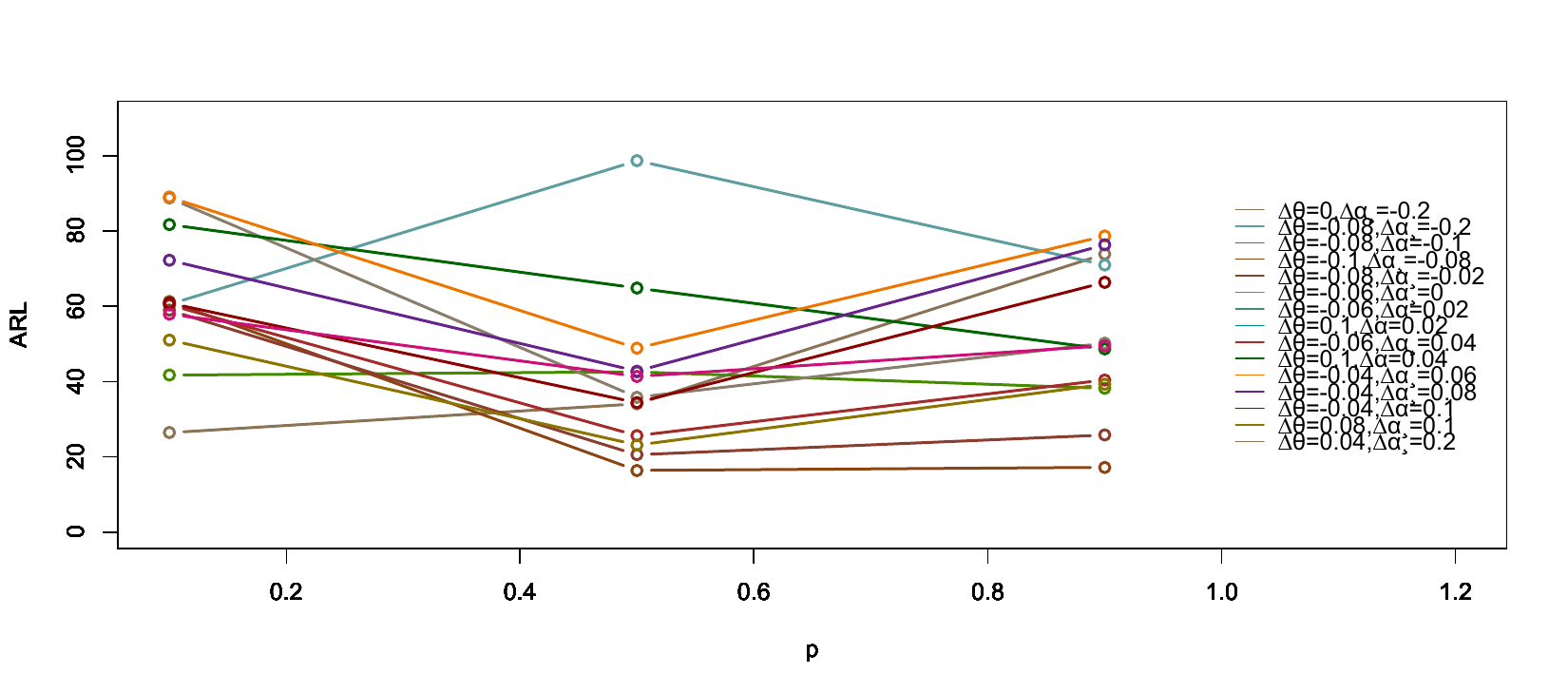}
\centering{$\left(e\right)$ $ARL_1$ for different choices of $p$}
\end{minipage}
\quad
\begin{minipage}[b]{0.48\linewidth}
\includegraphics[height=4.2 cm, width=7.2 cm]{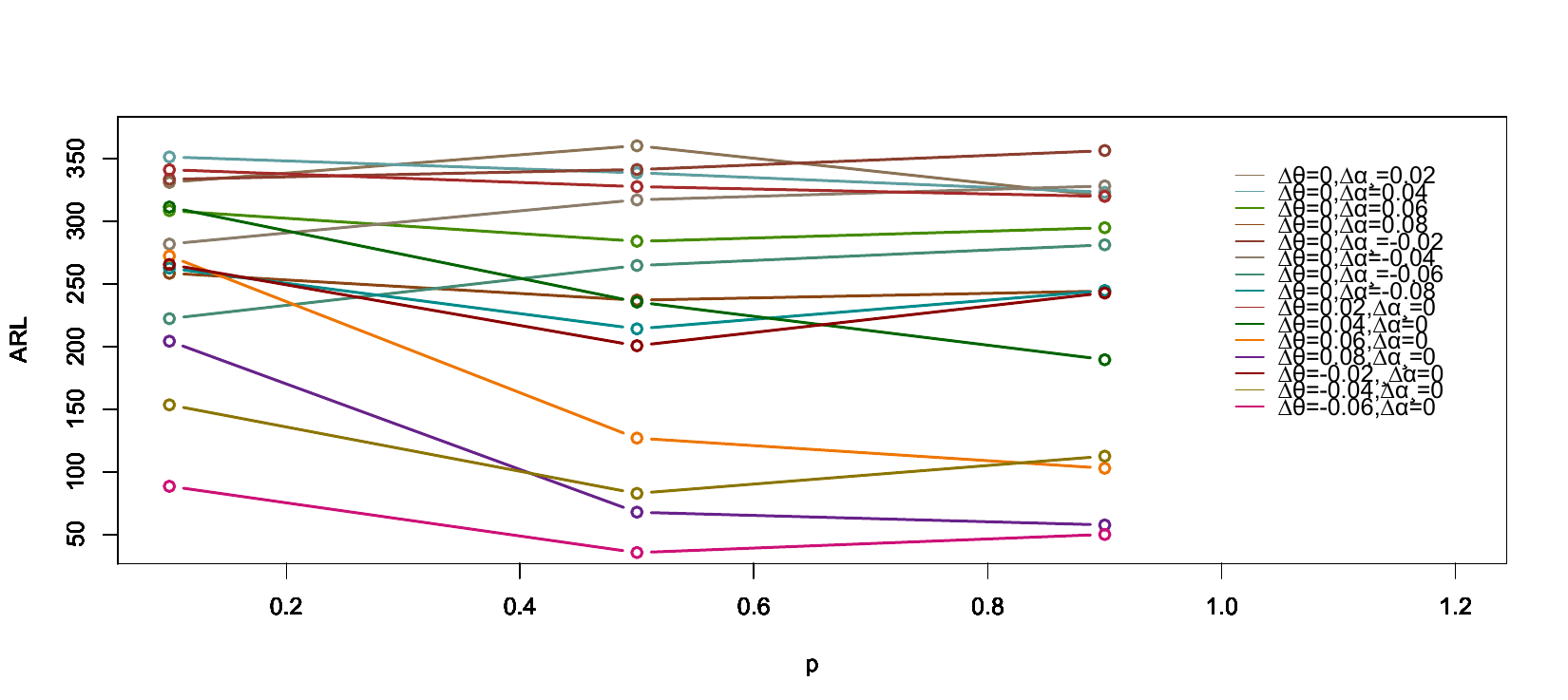}
\centering{$\left(f\right)$ $ARL_1$ for different choices of $p$}
\end{minipage}\caption{\label{figure1} Graphs of $ARL_1$ for different choices of $\Delta\theta$, $\Delta\alpha$ and $p$}
\end{figure}


\begin{thebibliography}{99}
\bibitem{ba} Balakrishnan N and Kundu D. Hybrid censoring: Models, inferential results and applications. {\it Computational Statistics \& Data Analysis.} 2013; 57(1): 166-209.
\bibitem {chi1} Chiang JY, Jiang N, Brown TN, Tsai TR and Lio YL. Control charts for generalized exponential distribution percentiles. {\it Communications in Statistics-Simulation and Computation.} 2017; 46(10): 7827-7843.
\bibitem {chi2} Chiang JY, Lio YL, Ng HKT, Tsai TR and Li. Robust bootstrap control charts for percentiles based on model selection approaches. {\it Computers \& Industrial Engineering.} 2018; 123: 119-133.
\bibitem{sh} Chowdhury S, Kundu A, and Modok B. Bootstrap beta control chart for monitoring proportion data. {\it International Journal of Quality \& Reliability Management.} 2021; 39(10): 2354-2377.
\bibitem {ef} Efron B and Tibshirani r J. An Introduction to the Bootstrap. {\it Chapman \& Hall.} 1993; New York.
\bibitem{ep} Epstein B. Truncated Life Tests in the Exponential Case. {\it The Annals of Mathematical Statistics.} 1954; 25(3): 555–564.
\bibitem{er} Erto P and Pallotta G. A new control chart for Weibull technological processes. {\it Quality Technology \& Quantitative Management.} 2007; 4(4): 553-567. 
\bibitem{er1} Erto P, Pallota G and Mastrangelo CM. A semi-empirical Bayesian chart to monitor Weibull percentiles. {\it Scandinavian Journal of Statistics.} 2015; 42(3): 701-712. 

\bibitem{ha} Haghighi F, Pascual F and Castagliola P. Conditional control charts for Weibull quantiles under type-II censoring. {\it Quality and Reliability Engineering International.} 2015; 31(8):1649-64.
\bibitem {hy} Hyndman RJ and Fan Y. Sample quantiles in statistical packages. {\it American Statistician.} 1996; 50: 361-365.
\bibitem {jon} Jones LA and Woodall WY. The performance of bootstrap control charts. {\it Journal of Quality Technology.} 1998; 30: 362-375.
\bibitem{lee} Lee E and Wang J. Statistical methods for survival data analysis. {\it John Wiley and Sons, New York.} 2003;
\bibitem {li} Lio Y L and Park C. A bootstrap control chart for Birnbaum-Saunders percentiles. {\it Quality and Reliability Engineering International.} 2008; 24: 585-600. 
\bibitem {li1} Lio Y L and Park C. A bootstrap control chart for inverse Gaussian percentiles. {\it Journal of Statistical Computation and Simulation.} 2010; 80: 287-299.
\bibitem {li2} Lio YL, Tsai TR, Aslam M and Jiang N. Control charts for monitoring Burr type-X percentiles. {\it Communications in Statistics-Simulation and Computation.} 2014; 43: 761-776.

\bibitem {li3} Liu RY and Tang J. Control charts for dependent and independent measurements based on the bootstrap. {\it Journal of the American Statistical Association.} 1996; 91: 1694-1700.
\bibitem{lu} Louis A. Finding the observed information matrix using the EM algorithm. {\it Journal of the Royal Statistical Society.} 1982; 44: 226-233.
	\bibitem {ms1} Mudholkar G.S. and Srivastava D.K. Exponentiated Weibull family for analyzing bathtub failure-rate data. {\it IEEE Transactions on Reliability.} 1993; 42: 299-302.
	\bibitem {ms2} Mudholkar G.S. Srivastava D.K. and Friemer M. The exponential Weibull family: a re-analysis of the bus-motor-failure data. {\it Technometrics.} 1995; 37: 436-445.
	\bibitem {ms3} Mudholkar G.S. Srivastava D.K. and Kollia G.D. A generalization of the Weibull distribution with applications to the analysis of survival data. {\it Journal of the American Statistical Association.} 1996; 91: 1575-1583.
	\bibitem {na} Nadarajah S, Cordeiro G. M., and Ortega E. M. The exponentiated Weibull distribution: a survey. {\it Statistical Papers.} 2013; 54(3): 839-877.
	\bibitem{ng} Ng H.K.T. Chan P.S. Balakrishnan N. Estimation of parameters from progressively censored data using EM algorithm. {\it Computational Statistics and Data Analysis.} 2002; 39: 371-386.
	\bibitem {ni} Nichols MD and Padgett WJ. A bootstrap control chart for Weibull percentiles. {\it Quality and Reliability Engineering International.} 2005; 22: 141-151.
\bibitem {pa} Padgett WJ and Spurrier JD. Shewhart-type charts for percentiles of strength distributions. {\it Journal of Quality Technology.} 1990; 22: 283-288.
\bibitem {se} Seppala Moskowitz H, Plante r and Tang J. Statistical process control via the subgroup bootstrap. {\it Journal of Quality Technology.} 1996; 27: 139-153. 
\bibitem{vi} Vining G, Kuahci M and Pedersen S. Recent advances and future directions for quality engineering. {\it Quality and Reliability Engineering International.} 2016; 32(3): 863-875.
\bibitem{wa} Wang FK, Bizuneh B, and Cheng XB. New control charts for monitoring the Weibull percentiles under complete data and Type-II censoring. {\it Quality and Reliability Engineering International.} 2018; 34(3): 403-16.
\end{thebibliography}
\end{document}